\documentclass[10pt,twocolumn,twoside]{IEEEtran}
\usepackage{amsmath,amsfonts}
\usepackage{algorithmic}
\usepackage{algorithm}
\usepackage{array}
\usepackage[caption=false,font=normalsize,labelfont=sf,textfont=sf]{subfig}
\usepackage{textcomp}
\usepackage{stfloats}
\usepackage{url}
\usepackage{verbatim}
\usepackage{graphicx}
\usepackage{cite}
\usepackage{caption}
\usepackage{tabularx}
\usepackage{hyperref}

\usepackage[switch]{lineno} % to be deleted
\begin{document}

%https://www.overleaf.com/project/63aa97c9dea1db2f10efd07b
\title{MuSE-SVS: Multi-Singer Emotional Singing Voice Synthesizer that Controls Emotional Intensity}

%\author{IEEE Publication Technology,~\IEEEmembership{Staff,~IEEE,}
\author{Sungjae Kim, Yewon Kim, Jewoo Jun, and Injung Kim\footnote{Corresponding author.}
        % <-this % stops a space
%\thanks{This paper was produced by the IEEE Publication Technology Group. They are in Piscataway, NJ.}% <-this % stops a space
%\thanks{Manuscript received Feburary 11, 2023}
\thanks{The authors are with Computer Science and Electronical Engineering Department, Handong Global University, Pohang 37554, Korea. (e-mail: sjkim@handong.ac.kr; 21800132@handong.ac.kr; junfa0118@handong.ac.kr; ijkim@handong.ac.kr)}
}

% The paper headers
\markboth{Journal of \LaTeX\ Class Files,~Vol.~0, No.~0, Feburary~2023}%
{Shell \MakeLowercase{\textit{et al.}}: A Sample Article Using IEEEtran.cls for IEEE Journals}

\IEEEpubid{0000--0000/00\$00.00~\copyright~2023 IEEE}
% Remember, if you use this you must call \IEEEpubidadjcol in the second
% column for its text to clear the IEEEpubid mark.

\maketitle

\begin{abstract}
We propose a multi-singer emotional singing voice synthesizer, Muse-SVS, that expresses emotion at various intensity levels by controlling subtle changes in pitch, energy, and phoneme duration while accurately following the score. To control multiple style attributes while avoiding loss of fidelity and expressiveness due to interference between attributes, Muse-SVS represents all attributes and their relations together by a joint embedding in a unified embedding space. Muse-SVS can express emotional intensity levels not included in the training data through embedding interpolation and extrapolation. We also propose a statistical pitch predictor to express pitch variance according to emotional intensity, and a context-aware residual duration predictor to prevent the accumulation of variances in phoneme duration, which is crucial for synchronization with instrumental parts. In addition, we propose a novel ASPP-Transformer, which combines atrous spatial pyramid pooling (ASPP) and Transformer, to improve fidelity and expressiveness by referring to broad contexts. In experiments, Muse-SVS exhibited improved fidelity, expressiveness, and synchronization performance compared with baseline models. The visualization results show that Muse-SVS effectively express the variance in pitch, energy, and phoneme duration according to emotional intensity. To the best of our knowledge, Muse-SVS is the first neural SVS capable of controlling emotional intensity.

%We propose Muse-SVS, the first multi-singer emotional singing voice synthesizer that expresses various levels of emotional intensity. During synthesizing singing voices according to the lyrics, pitch, and duration of the music score, Muse-SVS reflects singer characteristics and emotional intensity by adding variances in pitch, energy, and phoneme duration according to singer ID and emotional intensity. Representing all attributes by conditional residual embeddings in a single unified embedding space, Muse-SVS controls mutually correlated style attributes, minimizing interference. Additionally, we introduce such novel techniques as Statistical Pitch Predictor, Context-aware Residual Duration Predictor, and ASPP-Transformer to accurately predict and reflect multiple attributes according to singer ID and emotion. In particular, we apply emotion embedding interpolation and extrapolation techniques that lead the model to learn a linear embedding space and allow the model to express emotional intensity levels not included in the training data. In experiments, Muse-SVS exhibited higher scores in subjective evaluations on overall audio quality and model's expressiveness compared to comparative models. Furthermore, the results of several quantitative evaluations and visualization suggest that the proposed idea are effective in accurate prediction of pitch and duration according to singer ID and emotional intensities.
\end{abstract}


\section{Introduction}

\IEEEPARstart{A}{} singing voice synthesis (SVS) model is a generative model that produces singing voices from the lyrics, note pitch, and note duration of a music score. Similar to text-to-speech (TTS), SVS converts texts (lyrics) into speech signals. Moreover, SVS has an additional requirement of synthesizing voices according to the pitch and duration of the notes. In the past, most SVS systems were based on traditional approaches such as a concatenative method and a hidden Markov model (HMM) \cite{ConcatSVS1997, HMM-SVS2006, Vocaloid2007}. Recently, end-to-end neural SVS has been actively studied \cite{ATK2019,XiaoiceSing2020,HifiSinger2020,N-Singer2021,DiffSinger2022,VISinger2022}. Neural SVS is more flexible than conventional methods and can effectively express various singing styles.

Reflecting the singer's characteristics and emotions is important for synthesizing natural and expressive singing voices. In particular, the intensity of emotion, as well as the type, is crucial in conveying the feeling of a song. However, there are few studies on expressing emotions in SVS, and they are based on traditional methods such as HMM \cite{ParamSVS2010}. To the best of our knowledge, there is no prior neural SVS model that expresses emotions of varying intensities \cite{SVSSurvey2021}. Regarding TTS, many studies have been conducted to express emotional types \cite{UnsupEmotion2012, StyleFactor2017, GST2018, RefGST2018, CHiVE2019, Mellotron2020, EmoTTSRich2020}, but there are few studies on expressing emotional intensities \cite{EmoTTSRich2020, EmoTTS2021, EMOQ-TTS2022}. The main challenge in speaker/singer ID and emotion control is disentangling them from other attributes such as pitch and rhythm. In particular, expressing emotions in SVS is more challenging than in TTS as the singing voice should accurately follow the pitch and duration of notes.

Previous studies have shown that emotions in a singing voice are mainly expressed by subtle changes in a fundamental frequency (F0) contour, power envelopes, and spectral sequences \cite{ExtractF02002, EmoTTSThree2015, ExpressiveSVS2018}.
Additionally, \cite{EmoInSV2017} reported that the level of loudness, the change in loudness, and the variations in F0 have a significant effect on emotional expression. Consequently, to effectively express emotions in a singing voice, it is essential to precisely model the variances in pitch, energy, and phoneme duration according to the type and intensity of the emotions. However, it is challenging to synthesize such variances while maintaining accurate pronunciation and following the pitch and duration of notes. Multi-singer emotional SVS is even more challenging because the model should express the timbre of multiple singers in addition to the aforementioned attributes.

\IEEEpubidadjcol
In this paper, we propose Muse-SVS, the first multi-singer emotional neural SVS model that effectively expresses the type and intensity of emotions. Muse-SVS synthesizes spectrograms in a non-autoregressive manner. Muse-SVS learns each style attribute by a residual embedding conditional on the preceding attributes to reflect the correlation between style attributes, thereby avoiding interference between them. More importantly, it predicts subtle variances in pitch, energy, and phoneme duration according to the emotions while accurately following the pitch and duration of notes. Muse-SVS learns a continuous emotion embedding space and can express emotional intensities not included in the training data, including stronger emotions than those in the training data, by applying emotion embedding interpolation and extrapolation. We also propose a novel ASPP-Transformer that combines atrous spatial pyramid pooling (ASPP) \cite{Deeplabv22017} and Transformer \cite{Transformer2017}. The ASPP-Transformer incorporates broad contexts while focusing on local details, thereby improving the fidelity of singing voices with a large variation in phoneme duration.

To train and evaluate Muse-SVS, we collected 12.32 hours of Korean singing voices, including the voice of four singers and seven emotional intensity levels: neutral, and three levels each of happiness and sadness. In MOS tests and quantitative evaluations, Muse-SVS exhibited improved voice quality, expressiveness, and synchronization performance compared to baseline models. Furthermore, the visualization results presented in Section \ref{subsubsec:visualization_results} suggest that Muse-SVS learns attribute embeddings highly correlated to singer ID and emotional intensity and also suggest that Muse-SVS is able to express emotions by controlling subtle changes in pitch, energy, and phoneme duration. The audio samples are available online \footnote{\href{https://muse-svs.github.io/}{https://muse-svs.github.io/}}. %\href{https://muse-svs.github.io/}{https://muse-svs.github.io/}.

%the proposed statistical pitch predictor produces subtle variances in F0 contour according to emotional intensity.

Our main contributions include: 1) the first neural multi-singer emotional SVS that expresses emotional type and intensity, 2) per-attribute residual encoders for expressive SVS that adds variance to phoneme embeddings while minimizing interference between interdependent and overlapping attributes, 3) statistical pitch predictor predicting pitch variance according to emotional intensity while reliably following the note pitch sequence, 4) context-aware residual duration predictor that prevents the accumulation of variances in phoneme duration for accurate synchronization with the score, 5) emotion embedding interpolation and extrapolation to learn a continuous emotional space, thereby expressing emotional intensities not included in the training data, and 6) ASPP-Transformer to incorporate broad contexts while focusing on local details of singing voices with minimal increase in computation and parameters.

\section{Related work}
\subsection{Single-singer SVS}
The SVS models developed in the early days of deep learning have similar structures to conventional SVS systems, but some modules have been replaced with deep neural networks \cite{FirstDNNSVS2016}. \cite{ATK2019} proposed an end-to-end neural SVS model composed of an encoder and an attention-based autoregressive decoder. In addition, they improved voice quality by applying adversarial loss in training. Regarding speech synthesis, \cite{FastSpeech2019} proposed a non-autoregressive TTS model, FastSpeech, that resolves the skipping and repeating issues in autoregressive models. Inspired by FastSpeech, many SVS models have adopted the non-autoregressive architecture based on the feed-forward Transformer (FFT) \cite{XiaoiceSing2020, HifiSinger2020}. They are composed of an encoder that extracts high-level embeddings from the lyrics, note pitch, and note duration, a length regulator that aligns a sequence of phoneme embeddings to the spectrogram frames, and a decoder that synthesizes spectrograms from the aligned phoneme embeddings. In addition, some SVS models are based on GAN \cite{WGansing2019, GANSVS2019, AnyoneGanSing2021} and diffusion model \cite{DiffSinger2022}.

\subsection{Multi-Singer SVS}
In the speech synthesis field, many multi-speaker TTS models reflect speaker ID by feeding a fixed-size speaker embedding into the decoder \cite{Deepvoicev22017, MultiSpeech2020}. Most multi-singer SVS models represent singer characteristics in similar ways \cite{WGansing2019, DurianSC2020, MR-SVS2022}. However, learning the characteristics of multiple singers requires sufficient data for each singer. A few studies have learned singers' characteristics from a small number of samples. \cite{VoiceCloningSVS2019} learned the characteristics of each singer by adapting a pre-trained SVS model to the target singer. \cite{DurianSC2020, MR-SVS2022, LSTandDualPitch2022} presented zero-shot style adaptation methods that apply a reference encoder to extract singer embeddings from reference audio samples. In particular, \cite{MR-SVS2022} and \cite{LSTandDualPitch2022} predicted frame-level singer embedding to capture time-varying singer characteristics by using an attention mechanism. Additionally, \cite{MR-SVS2022} applied multiple reference encoders to capture sufficient timbre information from multiple reference audio samples, while \cite{LSTandDualPitch2022} proposed a local style token module to control musical expression not specified in the music score.

\subsection{Emotion modeling in SVS}
Most previous studies on emotional expression in SVS are based on traditional approaches. \cite{ExtractF02002} analyzed the effect of F0 variances on emotional expression in singing voices, and proposed an F0 control module to reflect four types of F0 dynamics characteristics: overshoot, vibrato, preparation, and fine fluctuation. \cite{ParamSVS2010} proposed an SVS system based on hidden semi-Markov models (HSMM) that controls acoustic parameters affecting emotional expression. \cite{EmoTTSThree2015,ExpressiveSVS2018} analyzed the effect of F0 variance, amplitude envelope, and spectral sequence on emotional expression. \cite{EmoInSV2017, VocalExpinSing2021} reported that variations in F0 and duration significantly affects recognized emotion. In TTS, many prior studies have controlled emotion through emotion embeddings \cite{EmotionEmbedding2021}. However, to the best of our knowledge, there is no prior end-to-end neural SVS model that controls the type and intensity of emotion \cite{SVSSurvey2021}.

\subsection{Multiple style modeling in unified embedding space}
\begin{figure}[ht!] %% Figure of Unified Embedding Space
\vspace{-0.1in}
%\vskip 0.1in
%\begin{center}
%\centerline{\includegraphics[width=6cm, height=4cm]{Figures/2_v2.png}}
%\centerline{\includegraphics[width=5.4cm, height=3.6cm]{figures/2_v2.png}}
\centerline{\includegraphics[width=0.7\columnwidth]{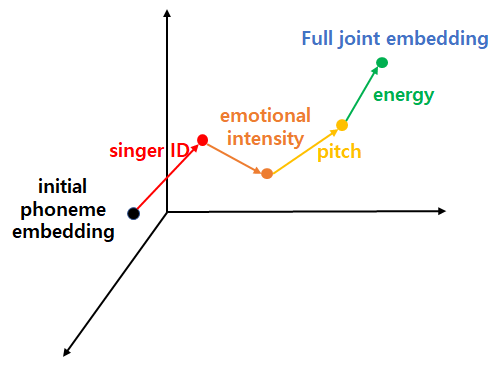}}
\caption[Conceptual illustration of unified embedding space]{The representation of multiple attributes by residual attribute embeddings in the unified embedding space. The coordinate of each point represents a joint embedding $E(y_i, z_{\leq k})$ of a phoneme $y_k$ combined with zero or more attributes, $z_1, ..., z_k$. The arrows represent the residual attribute embeddings $R(z_k|y_i,z_{<k})$ conditional on the previous attributes. The final coordinate corresponds to the full joint embedding $E(y_i, z_1, ..., z_N)$ that represents all style attributes and their relations combined with the phoneme.}
\label{figure:unified-embedding-space}
%\end{center}
%\vskip -0.2in
\vspace{-0.2in}
\end{figure}

Many previous studies have represented style attributes in separate embedding spaces, implicitly assuming that style attributes are independent \cite{MultiSpkEmoCNN2019, MultiSpkExpressiveTTS2022}. However, since style attributes are correlated in reality, controlling an attribute without considering its effect on other attributes can cause interference. This problem is particularly serious for the multi-singer emotional SVS that controls many attributes while following the note pitch and note duration. To minimize interference, previous studies decorrelate attribute embeddings by gradient reversal \cite{MultiSpkEmoTTS2021, CrossSpkEmoDisentangle2022} or represent the relation between attributes by hierarchical embeddings \cite{HierarchicalControllableTTS2018, HierarachicalExpressiveTTS2019, FullyHierarchicalFineGrained2020}. However, they are insufficient for style attributes that are inherently correlated in a non-hierarchical way. \cite{UniTTS2021} addressed this challenge by learning multiple attributes and their relations together in a unified embedding space as illustrated in Fig. \ref{figure:unified-embedding-space}.
%The multi-singer emotional SVS model should control multiple style attributes, such as singer ID, emotional intensity, pitch, energy, and rhythm. However, it is challenging to disentangle overlapping attributes, such as singer ID and emotion, which often interfere with each other \cite{UniTTS2021}. Moreover, since some attributes are affected by others, it is difficult to independently predict each attribute. 

We applied the idea of \cite{UniTTS2021} with modifications to fit SVS. Muse-SVS avoids interference between attributes by representing all attributes and their relations by a joint embedding $E(y_i, z_1, ..., z_N)$ in a unified embedding space, where $y_i$ is the $i$-th phoneme and $z_k$ denotes the $k$-th attribute. As shown in Fig. \ref{figure:unified-embedding-space}, Muse-SVS estimates the joint embedding by predicting the residual embedding $R(z_k|y_i,z_1,...,z_{k-1})$ of each attribute $z_k$ conditional on the previous attributes $z_1,...,z_{k-1}$, and then combining them with the initial phoneme embedding $E(y_i)$ as 
\begin{equation}
\begin{split}
E(y_i, z_1, ..., z_k) &= E(y_i,z_{<k}) + R(z_k|y_i,z_{<k}) \\ &= E(y_i)+\sum_{j=1}^k{R(z_j|y_i,z_{<j})}.
\end{split}
\end{equation}

\section{Muse-SVS} \label{subsection:muse-svs}

\begin{figure*}[!t]
\centering
    \subfloat[Overall Architecture]{\includegraphics[width=0.35\textwidth]{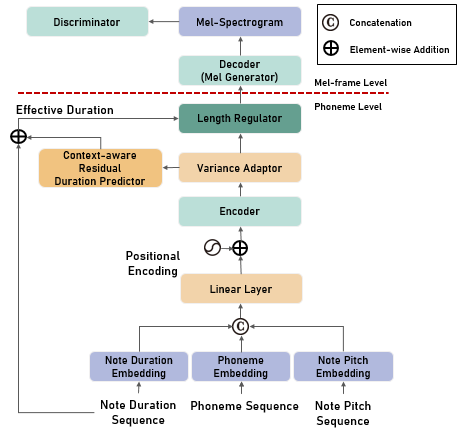}
    \label{figure:overall-architecture}}
    \subfloat[Variance Adaptor]{\includegraphics[width=0.35\textwidth]{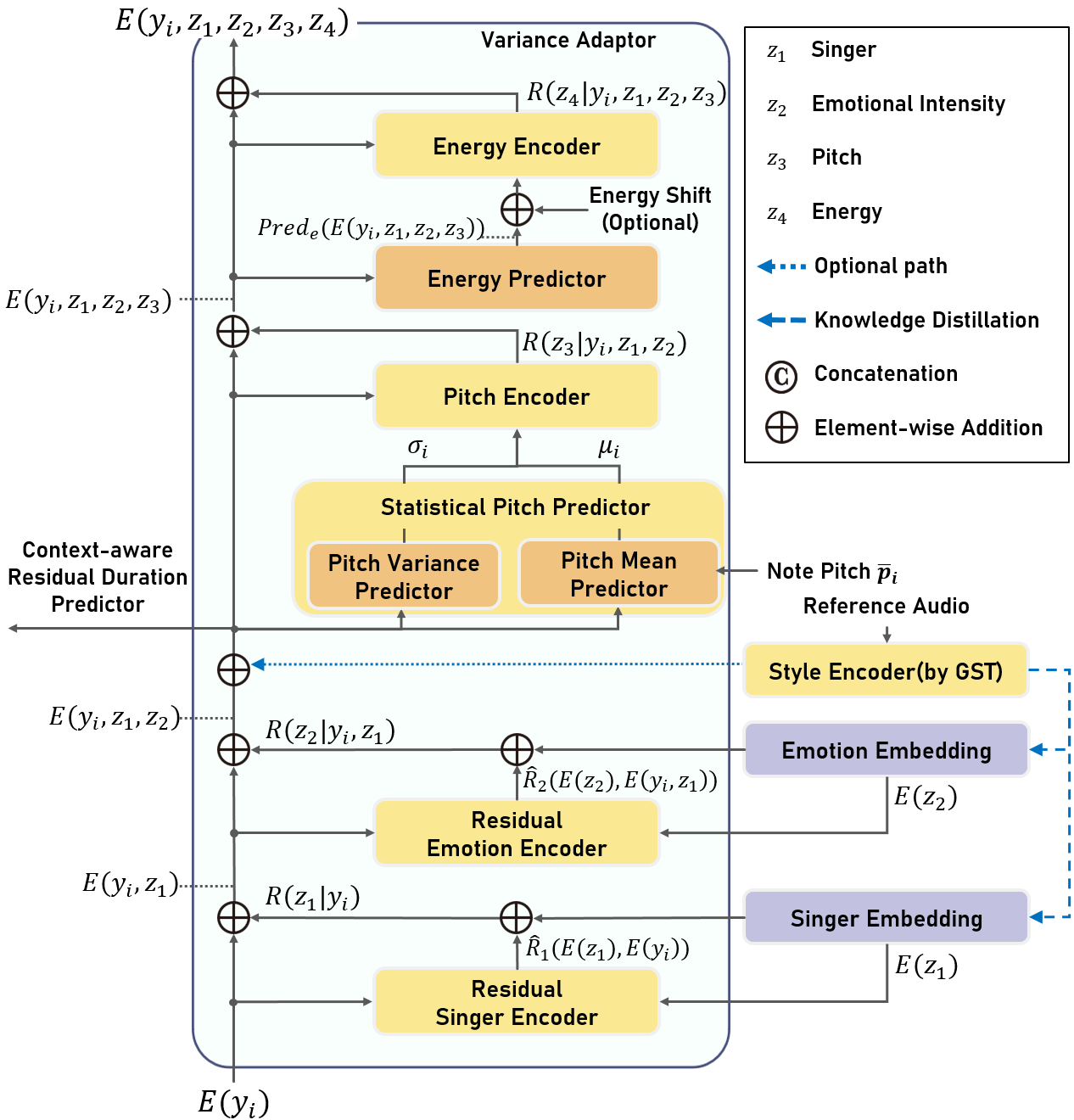}
    \label{figure:VA-architecture}}
    \hfil
    \subfloat[ASPP-Transformer]{\includegraphics[width=0.25\textwidth]{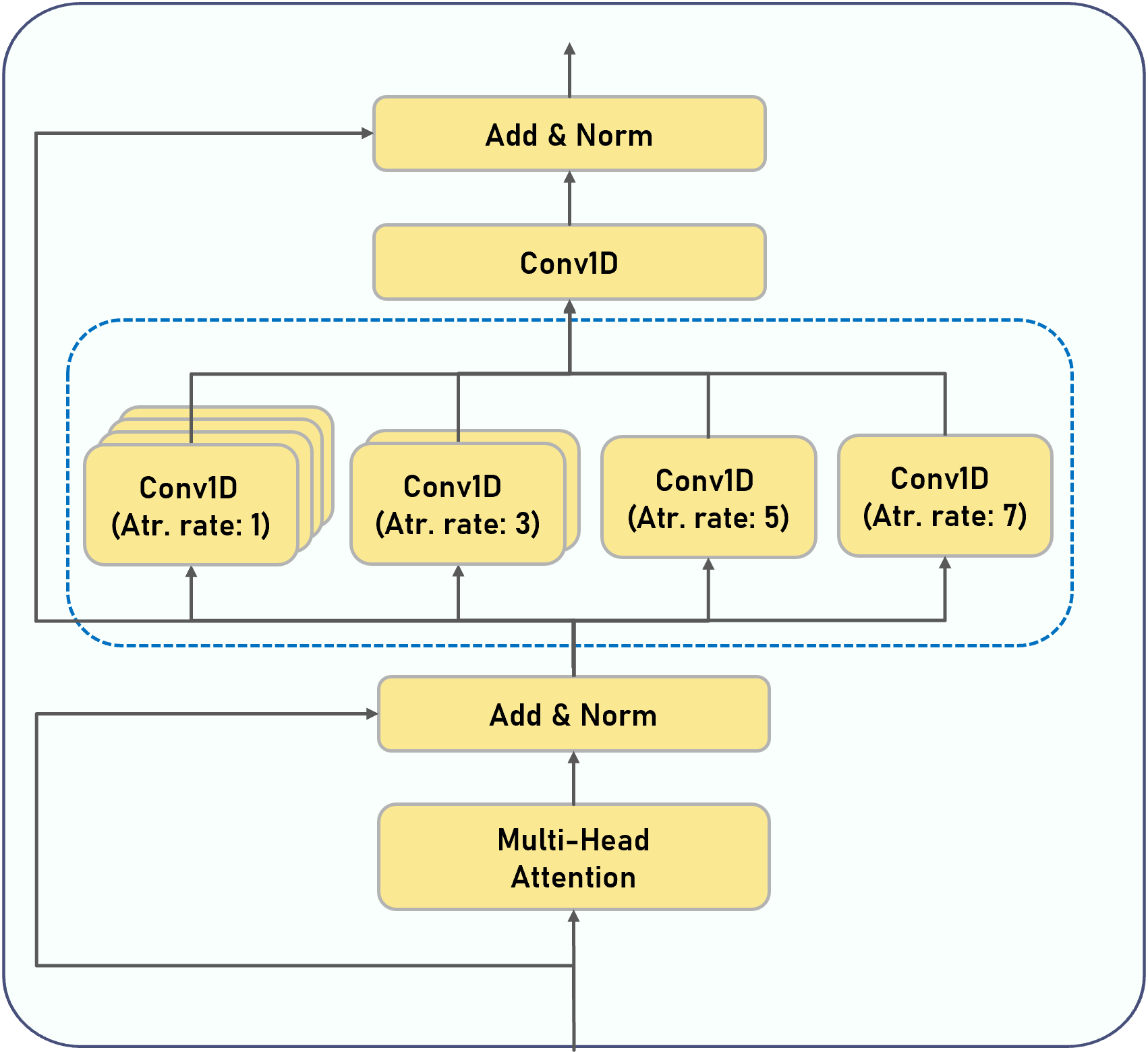}
    \label{figure:aspp-transformer}}
    %\captionsetup{justification=centering}
    \caption{The architecture of MuSE-SVS.}
\label{figure:architectures}
\vspace{-0.15in}
\end{figure*}

\subsection{Overall Architecture}
\label{subsection:overall architecture}

Fig. \ref{figure:overall-architecture} illustrates the structure of Muse-SVS. It takes lyrics (phoneme sequence), note pitch, and note duration as input. First, it converts the phonemes, note pitch, and note duration into embeddings using embedding tables. We concatenate the phoneme, pitch, and duration embeddings and apply a linear layer, in contrast to previous models \cite{XiaoiceSing2020}\cite{HifiSinger2020} that combine the embeddings by element-wise addition. The linear layer is a more general form than element-wise addition and exhibited higher fidelity in our preliminary experiments. Then, we add positional encoding \cite{Transformer2017} to the combined low-level embeddings. The encoder consists of FFT blocks and outputs the high-level representation of each phoneme combined with its pitch and duration information, which we call initial phoneme embedding and denote as $E(y_i)$.

% New version
% FastSpeech2, UniTTS 의 아류라는 관점으로 글이 읽히지 않도록 수정
% 1. Variance Adaptor 라는 명칭을 제거하고, Encoder-Decoder 구조로 설명
% 1-1. Encoder: joint embedding을 모델링
% 1-2. Decoder: joint-embedding -> Mel-Spectrogram

%좀 더 자세히 설명
%Muse-SVS consists of encoder-decoder architecture. Some attributes of a singing voice are affected by the others. Therefore, it is difficult to predict each attribute independently. Therefore our encoder models joint embedding $E(y_i,z_{1,\cdots,k})$.
%To predict joint embedding, we apply
%we designed the variance adaptor to predict style attributes in an auto-regressive manner, which predicts the style attribute given the previous state. $E(y_i,z_{1,\cdots,k}$=$E(y_i)+E(z_1|y_i)$

The variance adaptor adds variance information to the initial phoneme embedding $E(y_i)$ according to style attributes $z_k$ to build the full joint embedding $E(y_i, z_1,...,z_N)$, as shown in Fig. \ref{figure:VA-architecture}. The variance adaptor consists of a collection of per-attribute predictors and encoders. Each attribute predictor predicts the value of the corresponding attribute, such as pitch and energy, while the attribute encoder produces the residual embedding $R(z_k|y_i,z_{<k})$ of the attribute. The variance adaptor adds the residual embeddings of style attributes $z_1, ..., z_N$ to the phoneme embedding sequentially, where $z_1,...,z_4$ are singer ID, emotional intensity, pitch, and energy, respectively. While the variance adapter adds attributes to the phoneme, the embedding moves along the path $E(y_i)$, $E(y_i, z_1)$, $E(y_i, z_1, z_2)$, ..., $E(y_i, z_1, ..., z_N)$, where $E(y_i, z_{\leq k}) = E(y_i, z_{<k}) + R(z_k|y_i,z_{<k})$ is a joint embedding that represents the phoneme $y_i$ combined with attributes $z_1, ..., z_k$. In Fig. \ref{figure:VA-architecture}, the vertical arrows correspond to the joint embeddings $E(y_i, z_{\leq k})$ accumulating the residual attribute embeddings $R(z_k|y_i,z_{<k})$ for $1 \leq k \leq 4$. % Since the residual embedding is conditional on the previous attributes, it represents the attribute embeddings adapted to the preceding attribute. 

%Muse-SVS represents each attribute by a residual embedding, which is the vector distance between the embeddings before and after applying the attribute, i.e., $R(z_k|y_i, z_{<k}) = E(y_i, z_{\leq k}) - E(y_i, z_{\leq k-1})$. 

Muse-SVS learns the residual attribute embeddings with residual attribute encoders $\hat{R}_k(\cdot)$ that take as input the previous joint embeddings $E(y_i, z_{<k})$ to reflect the phoneme and the previous attributes. In Fig. \ref{figure:VA-architecture}, the horizontal arrows from the vertical arrows represent the joint embeddings transmitted to the attribute predictors and encoders. Additionally, the singer and emotion encoders take as input attribute embeddings $E(z_k)$ retrieved from singer and emotion embedding tables, respectively.
The embedding table learns the mean embedding of attribute $z_k$, in which the influence of the previous attribute was removed by normalization \cite{UniTTS2021, RobustFineProsody2019}. The residual encoder $\hat{R}_k(E(z_k), E(y_i, z_{<k}))$ estimates the shift from $E(z_k)$ for adapting to the phoneme $y_i$ and for reflecting the influence of the previous attributes. Therefore, the singer and emotion encoders have the form $\hat{R}_k(E(y_i, z_{<k}), E(z_k))$. The residual singer and emotion embedding tables were learned through knowledge distillation from a style encoder that is based on the global style token (GST) \cite{GST2018}. For more details, refer to \cite{UniTTS2021}. 

On the other hand, the pitch ($z_3$) and energy ($z_4$) encoders take as input the output of the corresponding predictor, as $\hat{R}_k(E(y_i, z_{<k}), Pred_k(E(y_i,z_{<k}))$, where $Pred_k(\cdot)$ is either the pitch predictor $Pred_p(\cdot)$ or the energy predictor $Pred_e(\cdot)$. As the pitch and energy predictors refer to the previous attributes through the joint embedding $E(y_i,z_{<k})$, they predict pitch and energy differently according to emotional intensity, as shown in Fig. \ref{figure:f0-vi-woutSPP-ours} and Fig. \ref{figure:energy-contours}.

The duration predictor predicts the duration of each phoneme. We use $E(y_i, z_1, z_2)$ as its input because the duration can be affected by $z_1$ (singer ID) and $z_2$ (emotion), but independent of $z_3$ (pitch) and $z_4$ (energy). Consequently, it predicts phoneme durations differently according to emotional intensity, as shown in Fig.\ref{figure:mel-spectrogram}.
The length regulator aligns the sequence of joint embeddings $E(y_i,z_{1,...,4})$ to Mel-frames by duplicating the joint embeddings for the duration of each phoneme, similar to \cite{FastSpeech22020}. The decoder converts the aligned joint embeddings into a Mel-spectrogram. We also added a discriminator to improve fidelity through adversarial loss, similar to \cite{ATK2019, HifiSinger2020, N-Singer2021}.

\subsection{Statistical Pitch Predictor}

SVS synthesizes a singing voice that follows the note pitch sequence. In addition to the macroscopic trend that matches the note pitch sequence, the emotional SVS should generate microscopic changes within the interval of each note because emotional intensities are often expressed by subtle pitch changes such as vibrato. SVS models without emotion modeling produce F0 sequences as close to the training samples as possible at the frame level without separating the two types of pitch changes \cite{XiaoiceSing2020, HifiSinger2020, VISinger2022, Sinsy2021}. However, such an approach is sub-optimal for the emotional SVS because of a few reasons. First, estimating exact F0 trajectories is difficult because minute changes in the F0 are influenced by many situational factors \cite{Sinsy2021} and are not sufficiently consistent. Second, to express emotion, it is not necessary to reproduce the F0 trajectory perfectly, including the phase of microscopic fluctuation, and imitating local pitch variance is sufficient. Third, the emotional SVS should control local pitch variation while keeping the macroscopic trend close to the note pitch sequence. Therefore, we propose a novel statistical pitch predictor that estimates local pitch variances according to emotional intensity while reliably generating macroscopic pitch trends.

\begin{figure}[t!]
\begin{center}
\centerline{\includegraphics[width=9cm,height=7.5cm]{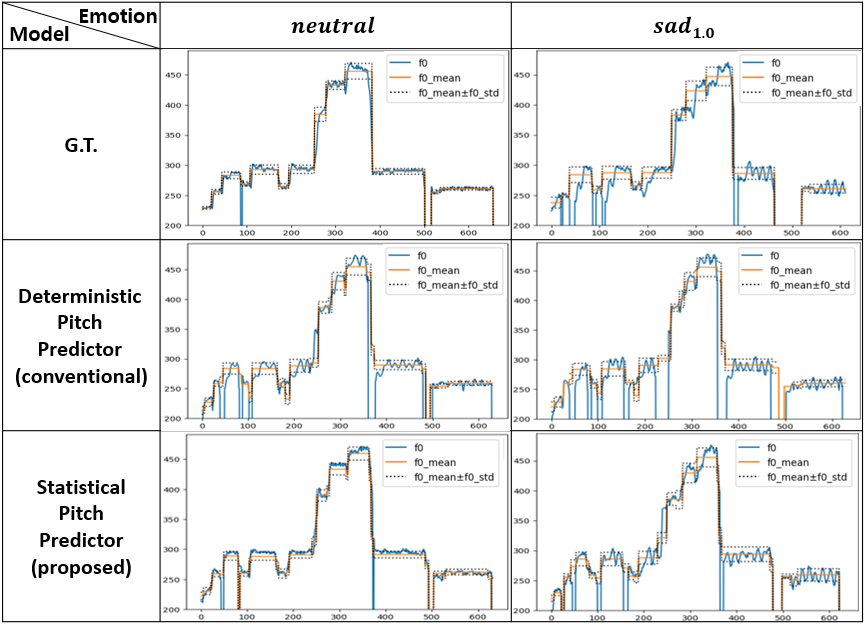}}
\caption[The F0 contours of the ground truth and synthesized singing voices]{The F0 contours (blue solid line) of a female singer's voices with emotional intensity levels $neutral$ (left) and $sad_{1.0} $(right). Each row displays the F0 contour of the ground truth samples (top), the output of Muse-SVS with the conventional deterministic pitch predictor (middle), and the output of Muse-SVS with the proposed statistical pitch predictor (bottom). The orange lines denote the mean of the F0 frequencies within the interval of each phoneme, and the black dotted lines denote the trajectory of mean$\pm$stdev. The deterministic pitch predictor produced vibrato with similar strengths regardless of emotional intensity. However, the proposed statistical pitch predictor produced vibrato of different strengths depending on emotional intensity while maintaining the macroscopic trend close to the note pitch sequence. }
\label{figure:f0_contour_mean_std}
\end{center}
\vskip -0.4in
\end{figure}

The proposed statistical pitch predictor estimates the distribution of the F0 frequencies at the phoneme level. The pitch predictor consists of a pitch mean predictor and a pitch variance predictor, and it estimates the local mean $\mu_i$ and local variance $\sigma_i^2$ of the F0 frequencies within the interval of each phoneme $y_i$. Both predictors take as input the joint embedding $E(y_i,z_1,z_2)$ to reflect the influence of singer ID ($z_1$) and emotional intensity ($z_2$). To reliably estimate mean pitch, the pitch mean predictor $Pred_{pm}(\cdot)$ estimates mean pitch $\hat{\mu}_i$ indirectly by first predicting the residual $\hat{r}_i$ between the note pitch $\bar{p}_i$ and the ground truth mean pitch $\mu_i$ measured from a training sample and then adding the predicted residual to the note pitch as $\hat{\mu}_i = \bar{p}_i + \hat{r}_i$. Previous studies have shown that such a residual pitch predictor helps mitigate the off-pitch problem \cite{XiaoiceSing2020, Sinsy2021}. The pitch variance predictor $Pred_{pcv}(\cdot)$ predicts the coefficient of variation $CV_i=\sigma_i / \mu_i$ instead of $\sigma_i^2$ because pitch variances tend to be correlated with pitch means, whereas in $CV_i$, the correlation with pitch means is removed by normalization, making $CV_i$ more reliably predictable \cite{EmoInSV2017}. During training, we optimize $Pred_{pm}(\cdot)$ and $Pred_{pcv}(\cdot)$ by the loss function presented in Eq. \ref{eq:pitch loss}, where $\mu_i$ and ${CV}_i$ are the ground truth mean pitch and pitch CV of phoneme $y_i$ and $\hat{CV}_{i}$ is the estimation of ${CV}_i$ predicted by $Pred_{pcv}(\cdot)$. $N_{pho}$ denotes the total number of phonemes. We set $\lambda_{pm} = 1$ and $\lambda_{pcv} = 10$.

\begin{equation}
\begin{split}
%    \mathcal{L}_{p} = \lambda_{pcv} \cdot \mathcal{L}_{pcv} + \lambda_{pm} \cdot \mathcal{L}_{pm}
    \mathcal{L}_{p} = & \lambda_{pm} \frac{1}{N_{pho}} \sum_i {\sqrt{((\bar{p}_i + \hat{r}_i) - \mu_i)^2}}  \\
                      & + \lambda_{pcv} \frac{1}{N_{pho}} \sum_i {\sqrt{(\hat{CV}_{i} - {CV}_i)^2}}
\end{split}    
\label{eq:pitch loss}
\end{equation}

%Fig. \ref{figure:f0_contour_mean_std} demonstrates the effectiveness of the proposed statistical pitch predictor. The left and right columns display the F0 contours of singing voices with emotional intensity levels $neutral$ (the lowest emotional level) and $sad_{1.0}$ (the strongest level of sadness), respectively. The top, middle, and bottom rows display the ground truth samples, the samples synthesized with the conventional deterministic pitch predictor and the samples synthesized with the proposed statistical pitch predictor, respectively. 

Fig. \ref{figure:f0_contour_mean_std} demonstrates the effectiveness of the proposed statistical pitch predictor. It displays the F0 contours of singing voices of the ground truth samples, the samples synthesized with the conventional deterministic pitch predictor, and the samples synthesized with the proposed statistical pitch predictor. In each F0 contour, the macroscopic trend follows the note pitch sequence, while the microscopic changes exhibit singing techniques to express emotions, such as vibrato and bending. In the ground truth samples, the strength of vibrato increases with emotional intensity. The deterministic pitch predictor estimated major trends similar to the note pitch sequences but failed to generate differences in microscopic variances according to emotional intensity. By contrast, the proposed statistical pitch predictor produced vibrato of different strengths depending on emotional intensity while maintaining the macroscopic trend close to the note pitch sequence. The results of an ablation study, shown in Table \ref{table:ablation-emotion-similarity}, also show that the statistic pitch predictor significantly improves the expressiveness of Muse-SVS.
 
%Since the statistical pitch predictor estimates a distribution, it cannot provide an exact estimate of F0 sequences. However, it helps to capture the emotion-affecting pitch patterns such as vibrato, as shown in Fig. \ref{figure:f0_contour_mean_std}. According to the result of the subjective evaluations(\ref{sec:subjective evaluation}), Muse-SVS exhibited higher MOS scores in terms of the similarity of emotional type and emotional intensity than the comparative models, and showed the effectiveness of the statistical pitch predictor in the ablation studies. The results suggest that predicting the distribution helps reflect various pitch patterns depending on emotional type, and emotional intensity.

\subsection{Context-aware Residual Duration Predictor (CRDP)}

%original
%In SVS, the accumulation of duration prediction error is fatal to maintain sync with instruments or other voice parts. However, it is challenging to predict accurate phoneme duration while controlling multiple attributes. To minimize duration prediction error, we utilize note duration by applying a context-aware residual duration predictor.

%Kor version
% 음소 길이의 오차 누적은 합성된 노래 음성과 함께 연주되는 악기와의 sync를 망가뜨릴 수 있기에, 노래 합성에 치명적인 문제이다. 하지만, 악보의 duration 을 따르면서, 가수와 감정에 따라 변화는 음소 길이의 미세한 변화를 반영하는 것은 상당히 어려운 문제이다. 이 문제를 해결하기 위해, 우리는 매 time-step 별로 노래 음성과 악기와의 sync를 고려하면서 각 음소 길이를 auto-regressive 하게 예측하는 CRDP (Context-aware Residual Duration Predictor) 를 개발하였다.

\begin{figure}[h!] %% Architecture of Context-aware Residual Duration Predictor
\begin{center}
\centerline{\includegraphics[width=\columnwidth, height=5cm]{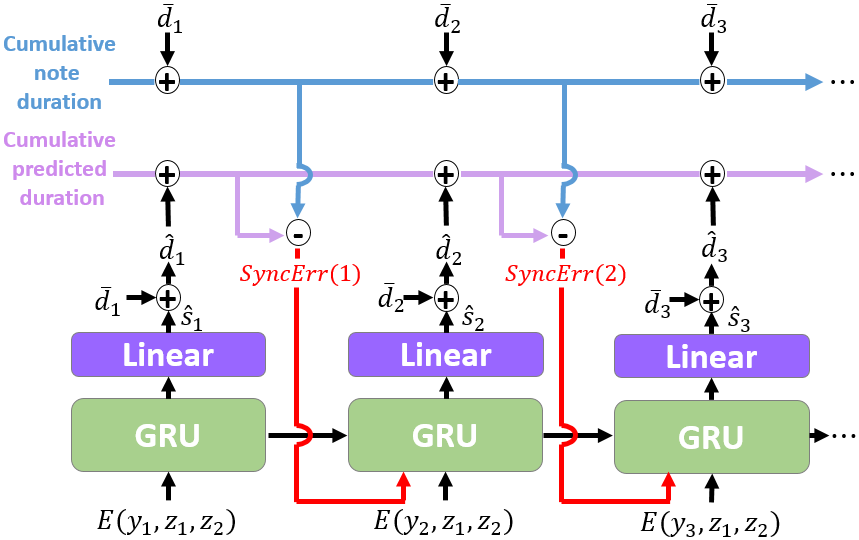}}
\caption{The architecture of context-aware duration predictor}
\label{figure:Context-aware-Residual-Duration-Predictor}
\end{center}
\vskip -0.2in
\end{figure}

Synchronization with the music score is one of the fundamental requirements in SVS. A major source of synchronization error is the difference in cumulative duration between the synthesized voice and the music score. However, for an emotional SVS, it is not straightforward to maintain synchronization. In the singing voice, emotions are often expressed by subtle variances in phoneme duration, and the emotional SVS should imitate such variances. Naively minimizing the difference between the predicted and note durations at the phoneme level can suppress intentional variances for emotional expression, thereby resulting in loss of expressiveness. However, the accumulation of such variances impairs the tempo of the synthesized singing voice, making it difficult to play with musical instruments. This issue is particularly significant when synthesizing a long singing voice. Consequently, the emotional SVS should synthesize subtle variations in the duration of individual phonemes while suppressing synchronization errors due to the accumulation of such variations.

In conventional SVS models, the duration predictor predicts the duration of all phonemes in one step with a parallel architecture \cite{XiaoiceSing2020, HifiSinger2020, VISinger2022}. However, it is hard to prevent the accumulation of phoneme-level variances with such parallel predictors because they predict the duration of each phoneme independently without considering contexts. To address this problem, we propose a novel context-aware duration predictor (CRDP) that minimizes synchronization errors by considering cumulative duration up to the previous phoneme when predicting the next phoneme duration while imitating variances in training samples to express emotions for individual phonemes. CRDP predicts the duration of phonemes sequentially with an autoregressive structure, as shown in Fig. \ref{figure:Context-aware-Residual-Duration-Predictor}. In predicting the duration of a phoneme, it takes the synchronization error of the previous phoneme and predicts the next phoneme duration inclined to compensate for the synchronization error.

At each step, CRDP takes as input the joint embedding $E(y_{i},z_{1},z_{2})$ to reflect the influence of singer ID and emotional intensity. CRDP also takes the synchronization error of the previous step $SyncErr(i-1)=\sum_{j=1}^{i-1}{\hat{d}_{j}} - \sum_{j=1}^{i-1}{\bar{d}_{j}}$ as input, where $\hat{d}_{j}$ and $\bar{d}_{j}$ denote the predicted and note durations of a phoneme $y_{j}$, respectively.
Similar to the pitch mean predictor, CRDP learns to estimate residual duration $s_i = d_i - \bar{d}_{i}$ as $\hat{s}_i = Pred_{d(i)}(E(y_i,z_1,z_2), SyncErr(i-1))$, where $d_i$ denotes the ground truth phoneme duration measured from the training sample. Then, CRDP adds the predicted residual to the note duration as ${\hat{d}_{i}} = {\bar{d}_{i}} + \hat{s}_i$.

It is noteworthy that the emotional SVS should learn from both the ground truth duration and the note duration. While the former is to learn emotional expression, the latter is to learn synchronization with the music score. During training, we optimize $Pred_{d(i)}(\cdot)$ using the duration loss presented in Eq. \ref{eq:duration loss}. The first term draws the predicted duration of the individual phoneme close to the ground truth duration, thereby leading the predictor to imitate the subtle variation in training samples to express emotions. The second term leads the predictor to learn to compensate for the synchronization error between cumulative predicted duration and cumulative note duration. Eq. \ref{eq:duration loss} does not penalize the difference between the predicted and note durations for individual phonemes but does penalize the difference in their accumulations. Consequently, CRDP maintains synchronization with the score without loss of expressiveness. In our experiments, we set $\lambda_{SyncErr} = 0.3$.

\begin{equation} \tag{3}
    \mathcal{L}_{d} = \frac{1}{N_{pho}} [\sum_i(\hat{d}_i - d_i)^2 + \lambda_{SyncErr} \cdot |\sum_{i}{SyncErr(i)}|]
    \label{eq:duration loss}
\end{equation}

\begin{figure}[h!] %% Figure of Comparison between Ours and XiaoiceSing
\vspace{-0.1in}
\begin{center}
\centerline{\includegraphics[width=\columnwidth, height=3.5cm]{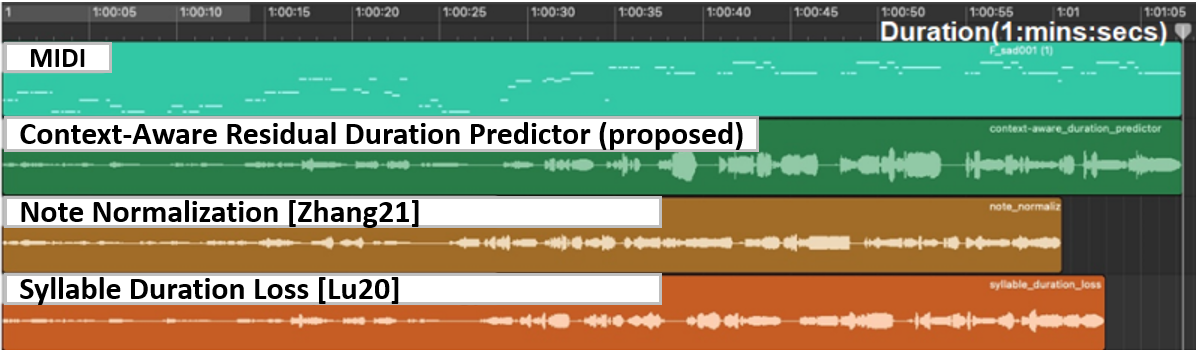}}
\caption[The synchronization errors of duration predictors for a song]{The synchronization errors of duration predictors for a long song composed of 117 notes whose length is 67.08 seconds. 1st row: the duration and pitch sequence of a song in MIDI format. 2nd-4th rows: singing voices synthesized with CRDP (proposed) and two baseline predictors applying note normalization \cite{VISinger2022} and syllable duration loss \cite{XiaoiceSing2020}, respectively. The note-level MAE of the three models are 0.064, 0.075, and 0.080 seconds from the top. However, their synchronization errors at the end of the song are 0.01, 6.96, and 4.83 seconds. CRDP exhibited a significantly lower synchronization error than the baseline predictors for a long song. }
\label{figure:Duration-error-comparison}
\end{center}
\vspace{-0.35in}
\end{figure}

% To CHECK: GT sequence의 note 수는?  - done
% To CHECK: 세가지 duration predictor 각각의 phoneme-level MAE는? - done
% => 기존 방법의 phoneme 별 duration 오차는 심각하지 않더라도 이들이 누적되면 심각해짐을 강조하기 위함 - done

Fig. \ref{figure:Duration-error-comparison} displays the duration and pitch sequence of a 67.07-second long song and the singing voices synthesized with CRDP and two baseline duration predictors. The first baseline predictor (3rd row) applies note normalization introduced by VISinger \cite{VISinger2022}, while the second baseline predictor (4th row) applies syllable duration loss used in XiaoiceSing \cite{XiaoiceSing2020}. The lengths of the audio samples are 67.07, 67.08, 60.11, and 62.24 seconds, respectively. The note-level MAE of the three models were merely 0.064, 0.075, and 0.080 seconds, implying that all three predictors reasonably estimate the duration of individual phonemes. However, the baseline models exhibited significant synchronization errors of 6.69 and 4.83 seconds at the end of the song, suggesting that minimizing prediction errors for individual phonemes is insufficient to prevent the accumulation of variances in phoneme duration. By contrast, the synchronization error of CRDP was 0.01 seconds, which is significantly lower than those of the baseline predictors. We also conducted a quantitative evaluation of synchronization performance and presented the result in Table \ref{table:synchronization error}.

%demonstrates the effectiveness of CRDP for a song that is 67.07 seconds long, compared with the method used in XiaoiceSing \cite{XiaoiceSing2020} and VISinger \cite{VISinger2022}. \cite{XiaoiceSing2020} improves prediction accuracy by applying syllable duration loss in addition to the phoneme duration loss. \cite{VISinger2022} predicts the ratio of the phoneme duration to the corresponding note duration and multiplies the ratio by note duration to compute the effective duration. The second row in the figure exhibits the duration estimated by CRDP, which is significantly more accurate than the third and fourth rows, which exhibit the duration predicted by the method of \cite{XiaoiceSing2020} and \cite{VISinger2022}.

\IEEEpubidadjcol
\subsection{Emotion Embedding Interpolation and Extrapolation}
Previous studies have reflected emotions in voices by feeding emotion embeddings to the model \cite{MultiSpkEmoTTS2021, EmoTTSRich2020, EmotionalStyleToken2019, UniTTS2021}. As described in  \ref{subsection:overall architecture}, Muse-SVS predicts residual emotion embeddings by combining an embedding table and a residual encoder as $R(z_2|y_i,z_1) = E_2(z_2) + \hat{R}_2(E_2(z_2), E(y_i, z_1))$. For simplicity, we denote the residual embedding for an emotional intensity level $v$ as $r_v = R(v|y_i,z_1)$.

\begin{figure}[ht] %% Figure of Emotion Embedding Interpolation and Extrapolation
%\vskip 0.1in
\begin{center}
\centerline{\includegraphics[width=\columnwidth]{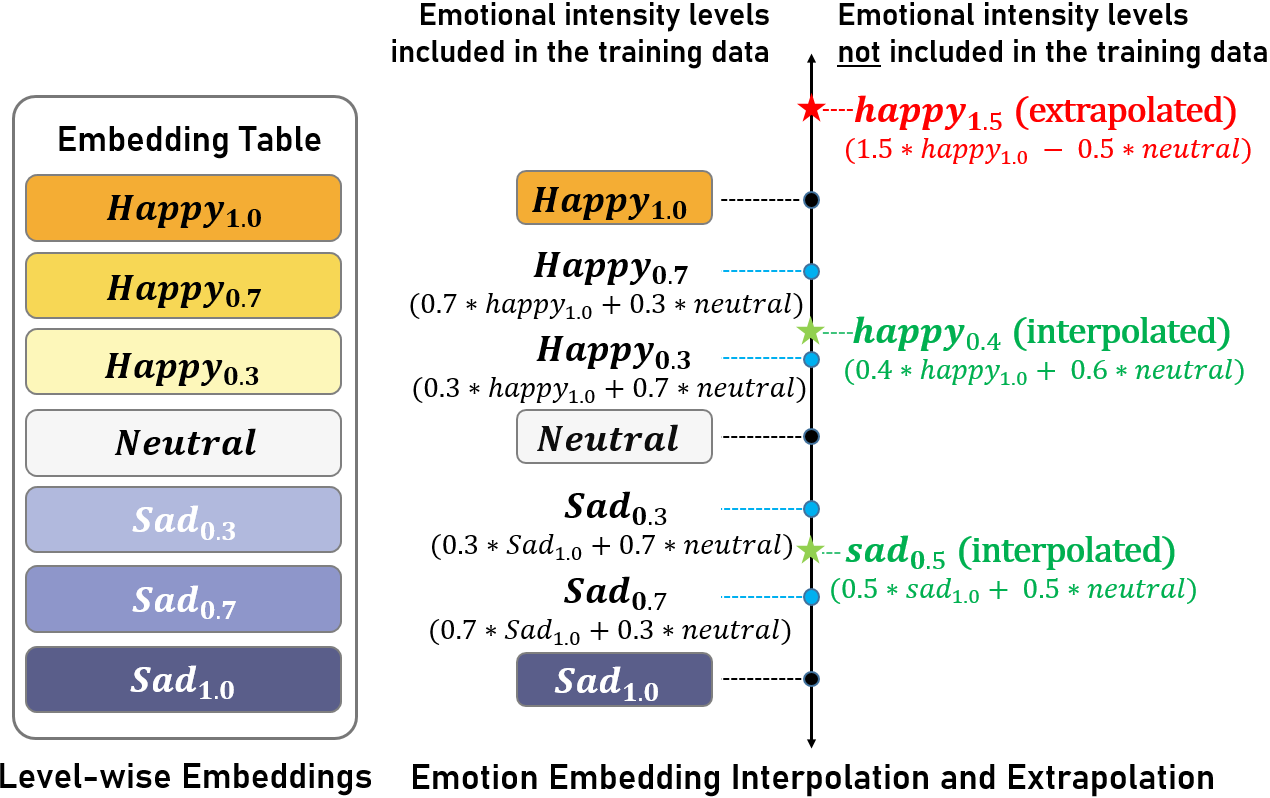}}
\caption[Emotion embedding]{The representation of emotional intensity levels by level-wise embeddings (left) and embedding interpolation (right). Muse-SVS can express emotional intensities that are not in the training data through emotion embedding interpolation and extrapolation.}
\label{figure:emotion-embedding-interpolation-extrapolation}
\end{center}
\vspace{-0.35in}
%\vskip -0.2in
\end{figure}

We considered two methods to learn the embeddings of multiple emotional intensity levels: level-wise embeddings, and embedding interpolation. The former learns separate embeddings for each emotional intensity level, while the latter learns only one embedding for each emotional type and represents intensity levels by embedding interpolation, as shown in Fig. \ref{figure:emotion-embedding-interpolation-extrapolation}.
Our training data contains seven emotional intensity levels: $happy_{1.0}$, $happy_{0.7}$, $happy_{0.3}$, $neutral$, $sad_{0.3}$, $sad_{0.7}$, and $sad_{1.0}$. Therefore, Muse-SVS learns seven separate embeddings with the level-wise embedding table. However, with emotion interpolation, it learns only three embeddings, each of which is for $happy_{1.0}$, $neutral$, and $sad_{1.0}$, and computes the embeddings of intermediate intensity levels by interpolation as $r_{happy_t}=t \cdot r_{happy_{1.0}} + (1-t) \cdot r_{neutral}$ and $r_{sad_t}=t \cdot r_{sad_{1.0}} + (1-t) \cdot r_{neutral}$. A prior study \cite{EmoTTSRich2020} also applies emotion interpolation to mix different types of emotion. However, they only apply embedding interpolation to synthesis, whereas we apply it to both training and synthesis.

Muse-SVS applies emotion embedding interpolation because it has multiple advantages over level-wise emotion embeddings. First, it is possible to express intermediate intensity levels, such as $happy_{0.5}$ and $sad_{0.8}$, that are not in the training data. Second, it enables the synthesis of singing voices with emotional intensities beyond those in the training data by emotion embedding extrapolation with $t > 1$. Our demo page presents audio samples produced by emotion embedding extrapolation. Third, applying emotion embedding interpolation during training causes the model to reflect the neighborhood relation between emotional intensity levels, thereby learning a linear embedding space, as shown in Fig. \ref{figure:emotion-embedding}.

\subsection{ASPP-Transformer}
The variations in phoneme duration in the singing voice is substantially higher than that of ordinary voices. To synthesize high-fidelity singing voices, the receptive field of the decoder should be sufficiently large because the decoder processes high-resolution feature maps at the frame level. Meanwhile, to effectively learn fine-grained acoustic features, which are important for expressing emotion \cite{EmoInSV2017}, the decoder should catch local details as well. Each FFT block consists of a multi-head self-attention (MSA) sublayer and a feed-forward sublayer composed of two convolution operators.

Previous studies have shown that convolution refers to a limited context \cite{NLN2018, ContextNet2020, GlobalNonLocal2021}. Although the MSA sublayer refers to broader contexts than convolution, enhancing the receptive field of convolution is essential because convolution plays an important role in Muse-SVS. In the preliminary experiments, we observed that the average activation of the feed-forward sublayers was twice higher than that of the MSA sublayers, suggesting that the contribution of convolution can be greater than MSA. However, simply enlarging convolution filters drastically increases computational cost and the number of parameters, thereby increasing the risk of overfitting.

To overcome this challenge, we extended FFT by replacing convolution with atrous spatial pyramid pooling (ASPP) \cite{Deeplabv22017}, as shown in Fig. \ref{figure:aspp-transformer}. This new building block is called ASPP-Transformer. ASPP-Transformer inherits the advantage of ASPP in that it can refer to a broad context, with minimal increase in computation and parameters. Focusing on the local neighborhoods while incorporating a broad context, we assigned a large number of channels to the filters with low atrous rates.

\begin{figure}[h!] %% Figure of Emotion Embedding Interpolation and Extrapolation
%\vskip 0.1in
\begin{center}
\centerline{\includegraphics[width=\columnwidth]{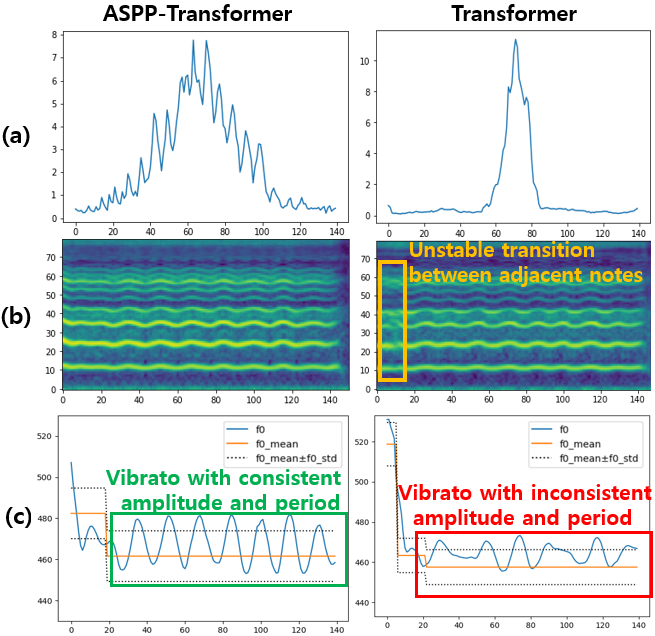}}
\caption{The effective receptive fields and output of the decoders composed of the proposed ASPP-Transformer and ordinary Transformer. The first row displays the $L_1$ norm of the gradient of the 70th output node with respect to the input embeddings aligned to frame-level resolution ($\frac{\partial x_{70}}{\partial E(y_j,z_{1:4})}, 1 \leq j \leq 140)$. The second and third rows display the synthesized Mel-spectrograms and their F0 contours, respectively.}
\label{figure:erf of ASPP-Transformer}
\end{center}
\vskip -0.4in
\end{figure}

Fig. \ref{figure:erf of ASPP-Transformer} compares the effective receptive fields of the decoders composed of the proposed ASPP-Transformer and ordinary Transformer blocks. The first row displays the gradient norm of an output node with respect to the input embeddings aligned to the frame-level resolution by the length regulator. The gradient norm measures the importance of the input on the output \cite{erf2016}. The right column shows that the ordinary Transformer has a limited effective receptive field, leading to discontinuous Mel-spectrogram and unstable vibrato. In contrast, ASPP-Transformer refers to broader contexts than ordinary Transformer, producing a more stable Mel-spectrogram and vibrato, thereby leading to improved fidelity and expressiveness.

\subsection{Training of Muse-SVS}
%Muse-SVS produces the Mel-spectrogram of a singing voice from lyrics, note pitch, and note duration, reflecting the specified singer ID, and emotional intensity. 
We train Muse-SVS by minimizing the $L_1$ reconstruction loss $\mathcal{L}_{m}$ between the ground truth and the synthesized Mel-spectrograms. The synthesized Mel-spectrogram often has a different length from the ground truth Mel-spectrogram. Therefore, we apply soft dynamic time warping \cite{SoftDTW2017} to align the two Mel-spectrograms. We also minimize the $L_2$ reconstruction loss between the predicted pitch, energy, and duration, and those of the training sample to improve the accuracy of the predictors. In addition, we combine an adversarial loss $\mathcal{L}_{adv}$ to alleviate the over-smoothing issue, following \cite{ATK2019, N-Singer2021, HifiSinger2020}. The total loss $\mathcal{L}_{total}$ combines those losses by weighted sum as Eq. \ref{eq:total_loss}, where $\mathcal{L}_{p}$, $\mathcal{L}_{e}$, and $\mathcal{L}_{d}$ are the $L_2$ losses for pitch, energy, and duration, respectively. $\lambda_{m}$, $\lambda_{p}$, $\lambda_{e}$, $\lambda_{d}$, and $\lambda_{adv}$ denote the weights of the loses. In experiments, we set $\lambda_{m},\lambda_{p},\lambda_{e},\lambda_{d}$ to 1, 1, 0.8, 0.8. We initially set $\lambda_{adv}$ to 0.01 to warm up and then gradually increase to 0.5.

% TO DO: 모든 sub-loss의 정의를 명시할 것. 현재는 duration loss만 정의되어 있음.
%        - pitch loss는 III-B, emotion encoder를 학습하기 위한 loss는 III-D에 설명하고 나머지는 III-F에 정의할 것.
%        - emotion encoder가 reconstruction loss를 최소화 함으로써 다른 모듈들과 같이 학습될 경우 III-F에 그렇게 설명할 것.

\begin{multline}
\mathcal{L}_{total} = \lambda_{m}\mathcal{L}_{m}+\lambda_{p}\mathcal{L}_{p}+\lambda_{e}\mathcal{L}_{e}+ \lambda_{d}\mathcal{L}_{d}+\lambda_{adv}\mathcal{L}_{adv} \tag{4}
\label{eq:total_loss}
\vspace{-0.3in}
\end{multline}

\subsection{Data collection}
To collect a singing voice dataset, we selected 69 Korean pop songs and categorized them into happy and sad songs. Subsequently, four singers (two male and two female) sang the selected songs four times at different emotional intensity levels: $happy_{1.0}$, $happy_{0.7}$, $happy_{0.3}$, and $neutral$ for the happy songs, and $sad_{1.0}$, $sad_{0.7}$, $sad_{0.3}$, and $neutral$ for the sad songs. Two of the singers are professionals, while the other two are amateurs with good singing skills.

The biggest challenges were setting singing guidelines for each emotional intensity level and leading the singers to follow the guidelines while singing. We consulted a vocal trainer to set the singing guidelines for each emotional intensity level. The singing guidelines consist of vocalization methods, the ratio of respiration and resonance, and singing techniques to express each emotional intensity level. Afterward, the professional singers sang the songs at each emotional intensity level according to the guidelines. We collected one hour of reference singing voice samples from the two professional singers. With the singing guidelines and the reference singing voices, we guided the amateur singers to sing at different emotional intensity levels. In this way, we collected additional 11.23 hours of singing voice samples from the two amateur singers.

\begin{table*}[t!] %% Table of Model Configuration
\caption{The hyperparameters of Muse-SVS}   
\label{table:model-configuration}
\begin{center}
\begin{sc}
\begin{tabular}{lcccr}
\hline
%\toprule
%\multicolumn{2}{c}{\textbf{Hyperparameter}}& \textbf{Muse-SVS} \vspace{0.1cm}\\
\textbf{network} & \textbf{hyperparameter} & \\
\hline\hline \vspace{0.1cm}
%\midrule
phoneme embedding dimension & & 384 \\
\hline
{encoder} & layers & 6 \\
 & MSA heads & 2 \\
 & MSA hidden dim. & 384 \\
 & 1st conv1D kernel size & 9 \\
 & 1st conv1D filters & 1536 \\
 & 2nd conv1d kernel size & 1 \\
 & 2nd conv1d filters & 384 \\
 & dropout rate & 0.2 \\
 \hline
{decoder} & layers & 6 \\ 
& MSA heads & 2 \\
& MSA hidden dim. & 384 \\
& ASPP kernel size & 9 \\
& ASPP kernel dilation rate & [1, 3, 5, 7] \\
& ASPP filters & [768, 384, 192, 192] \\
& conv1d kernel size & 1 \\
& conv1d filters & 384 \\
\hline
{per-attribute predictors of VA} & layers & 2 \\
 & conv1D kernel & 3 \\
 & conv1D filter size & 384 \\
 & dropout rate & 0.5 \\
\hline
{per-attribute encoders of VA} & conv1D kernel & 3 \\
 & conv1D filter size & 384 \\
 & dropout rate & 0.5 \\
\hline
{context-aware residual duration predictor} & hidden dim & 384\\
\hline
{reference encoder} & layers & 6 \\
& conv2D kernel & (3, 3) \\
& conv2D filter size & (32, 32, 64, 64, 128, 128) \\
& conv2D stride & (2, 2) \\
& hidden dim of GRU & 192 \\
\hline
{style token layer} & tokens & 10 \\
& token dim. & 48 \\
& attention hidden dim. & 384 \\
& attention heads & 8 \\
\hline

{discriminator} & layers & 3 \\
& conv2D kernel & (9, 9) \\
& conv2D filter size & (1, 64, 64, 64, 64, 64) \\
& conv2D stride & (1, 1) \vspace{0.1cm}\\
\hline\hline
\multicolumn{2}{c}{\textbf{Total Number of Parameters}} & 101M \\
\hline
%\bottomrule
\end{tabular}
\end{sc}
\end{center}
\vspace{-0.1 in}
\end{table*}

\section{Experiments}
\subsection{Dataset}
For experiments, we combined our dataset described in the previous section and 2.12 hours of Korean singing voices in the Children's Song Dataset (CSD) \cite{CSD2020}. The combined dataset consists of 7,672 singing voice samples that are 5-10 seconds long, sung by five singers. Of the total samples, we used 7,120 for training and 552 samples for testing. Since the samples in the CSD dataset do not have emotion labels, we labeled all of them as $neutral$.

\subsection{Details of SVS models}
\subsubsection{\textbf{Muse-SVS}}
%kor version
%Muse-SVS는 Fig. \ref{figure:overall-architecture} 처럼 복수의 모듈로 이루어진다. Encoder는 6개의 FFT block으로 구성되며, 각각의 FFT block은 하나의 MSA layer와 2개의 1-D convolution layer로 구성된다. Variance adaptor는 각각의 style attribute 를 모델링하는 per-attribute predictor와 per-attribute encoder로 구성된다. CRDP는 하나의 GRU layer로 구성된다. Decoder는 우리가 제안하는 6개의 ASPP-Transformer 로 이루어진다. Discriminator는 \cite{HifiSinger2020}에서 제안하는 SF-GAN 을 사용하여 개발하였다. 

Muse-SVS consists of multiple modules, as shown in Fig. \ref{figure:overall-architecture}. The encoder consists of six FFT blocks. The variance adaptor (VA) consists of per-attribute predictors and per-attribute encoders. The context-aware residual duration predictor consists of a gated recurrent unit (GRU) and a linear layer. The decoder consists of six ASPP-Transformer blocks. We built the discriminator based on SF-GAN \cite{HifiSinger2020} and the reference encoder based on the global style tokens (GST) \cite{GST2018}. The hyperparameters of the modules are presented in Table \ref{table:model-configuration}.

\subsubsection{\textbf{Baseline models for comparison}}
%kor version
% Muse-SVS 는 최초의 딥러닝 기반 multi-singer emotional SVS 모델이기 때문에, 적절한 베이스라인 모델이 없다. 따라서, 우리는 FastSpeech2와 VISinger를 확장하여 multi-singer emotional SVS 모델을 개발하였고, 이를 베이스라인 모델로 사용하였다. 먼저, FastSpeech2의 확장 모델인 'MSME-FFTSinger'는 다음의 세 가지 측면에서 FastSpeech2 와 다르다. 1) TTS 모델인 astSpeech2를 SVS 모델로 확장하기 위해 Encoder가 note pitch, note duration, 그리고 phoneme의 embedding 을 입력으로 받도록 함. 2) Single speaker 모델인 FastSpeech2가 가수와 감정을 조절할 수 있도록 decoder가 singer ID와 emotion을 condition으로 Mel-Spectrogram을 생성하도록 설계함. 이를 위해, singer embedding 과 emotion embedding을 lookup table로 부터 추출하고, decoder에 입력으로 들어가는 phoneme embedding에 더함. 3) MSME-FFTSinger는 frame 수준에서 pitch와 energy를 예측하는 FastSpeech2와 다르게, 음소 수준에서 pitch와 energy를 예측한다. 노래 발화는 일반 발화보다 더 복잡한 pitch와 energy 패턴을 보이기 때문에 frame 수준에서 pitch 와 energy를 정확하게 예측하는 것은 상당히 어렵다. 우리는 예비 실험에서 frame 수준에서 pitch와 energy를 예측하는 모델이 좋은 품질의 노래 음성을 만들어내는데 실패하는 것을 확인하였다. 따라서, MSME-FFTSinger의 pitch predictor와 energy predictor가 음소 수준에서 pitch와 energy를 예측하도록 설계하였다.

We built two multi-singer emotional SVS models to compare with Muse-SVS by extending FastSpeech2 \cite{FastSpeech22020} and VISinger \cite{VISinger2022}.\footnote{Since Muse-SVS is the first multi-singer emotional SVS model, there is no existing baseline model for comparison.} These are refered to as MSME-FFTSinger and MSME-VISinger. The structure of MSME-FFTSinger differs from FastSpeech2 in three aspects: 1) The encoder was extended to take as input the combined embedding of the phoneme, note pitch, and note duration to synthesize singing voices. 2) To reflect singer ID and emotion, MSME-FFTSinger retrieves the singer and emotion embeddings from embedding tables and adds them to the combined phoneme embedding. 3) MSME-FFTSinger predicts pitch and energy at the phoneme level, while FastSpeech2 predicts them at the frame level. This modification improved stability in our preliminary experiments.

% kor version
% 공개된 VISinger 오픈소스 코드가 없어서, VISinger의 베이스라인 모델인 VITS를 확장하여 직접 개발하였다. VISinger를 multi-singer emotional SVS 모델로 확장하기 위해서 VITS에서 사용한 speaker identity modeling 방법론을 사용하였으며, 이는 다음과 같다. 1) posterior encoder와 normalizing flow의 residual block에 singer와 emotion embedding을 더하는 'global conditioning'\cite{Wavenet2016} 을 사용함. 2) decoder의 입력인 latent variable 에 singer, emotion embedding을 더함.

To build MSME-VISinger, we first reproduced VISinger from the open-source of its baseline model, VITS \cite{VITS2021} \footnote{https://github.com/jaywalnut310/vits}, because we could not find a publicly available implementation of VISinger. Subsequently, we extended it to a multi-singer emotional SVS by applying the techniques introduced in \cite{VITS2021}. We extended the residual blocks of the posterior encoder and the normalizing flow to be conditional on singer and emotion embeddings, and added singer and emotion embeddings to the input of the decoder.
%To implement MSME-VISinger, we reproduced it by extending VITS \cite{VITS2021} \footnote{https://github.com/jaywalnut310/vits} which is the baseline model of VISinger. For extending VISinger to a multi-singer emotional SVS, we applied the method used in VITS. First, we applied global conditioning \cite{Wavenet2016} in the residual blocks of the posterior encoder and normalizing flow by adding singer and emotion embedding. Second, we added singer and emotion embedding to the input latent variables of decoder.

\subsubsection{\textbf{Models for ablation study}}
% kor version
% 우리는 ablation study를 위해 3 가지 모델을 개발하였다. 먼저, Statistical Pitch Predictor의 효용성을 검증하기 위해 Muse-SVS(w/o SPP)(Statistical Pitch Predictor)를 개발하였다. Muse-SVS(w/o SPP)는 Muse-SVS의 pitch variance predictor를 제거하여, pitch encoder가 pitch mean($\mu_i$)만을 condition 으로 pitch embedding $R(z_3|y_i,z_1,z_2)$을 생성하는 모델이다. 다음으로, ASPP-Transformer 의 효용성을 검증하기 위해 Muse-SVS(w/o ASPP)를 개발하였다. Muse-SVS(w/o ASPP)는 Muse-SVS 의 ASPP-Transformer 를 일반 FFT block으로 대체한 세팅이다. 마지막으로, 통합된 임베딩 공간에서 이전 스타일 요소를 condition 으로 다음 스타일 요소를 예측하는 방법론의 효용성을 검증하기 위해 Muse-SVS(uncond) 를 개발하였다. Muse-SVS(uncond)는 각각의 스타일 요소 임베딩$E(z_k)$을 독립적으로 예측하고, decoder는 phoneme embedding과 예측된 스타일 요소 임베딩들의 합을 입력으로 받아 Mel-Spectrogram을 생성한다.

We built three more SVS models derived from Muse-SVS for the ablation study. In the first model, Muse-SVS(w/o SPP), the statistical pitch predictor was replaced with the deterministic pitch predictor. In the second model, Muse-SVS(w/o ASPP), the ASPP-Transformer blocks were replaced with ordinary FFT blocks. Finally, the third model, Muse-SVS(uncond) does not apply the conditional attribute predictors and encoders, and estimates attribute embeddings unconditionally with embedding tables $E(z_k)$.

%We designed `Muse-SVS(w/o SPP)' to evaluate the effectiveness of the statistical pitch predictor described in Section 2. Muse-SVS(w/o SPP) is identical to Muse-SVS with the difference that the pitch encoder of Muse-SVS(w/o SPP) doesn't refer to F0 CV. Muse-SVS(w/o ASPP) was designed to test the effectiveness of the ASPP-Transformer block. Muse-SVS(w/o ASPP) has the same architecture as Muse-SVS, while it applies the standard FFT block. Finally, Muse-SVS(uncond) is designed to evaluate the effectiveness of the method of modeling multiple style attributes in the unified embedding space by conditional residual embedding. Muse-SVS(uncond) differs from Muse-SVS in one aspect where the attribute encoders of Muse-SVS(uncond) predict attribute embeddings $E(z_k)$ independently of the previous attribute, while those of Muse-SVS predict $R(z_k|y_i, z_{<k})$.

\subsection{Training of the SVS models} 
Muse-SVS, MSME-FFTSinger, and the models for the ablation studies were trained for two days on a single RTX-3090 GPU with 24 GB memory. We used the ADAM optimizer with a learning rate of 0.001, $\beta_1$=0.9, $\beta_2$=0.98, $\epsilon$=$10^{-9}$, and batch size of 8. In addition, we followed the same learning rate schedule of \cite{Transformer2017}. We trained MSME-VISinger for 14 days on two RTX-3090 GPUs and a batch size of 8 for each GPU. MSME-VISinger was trained using the AdamW optimizer where the initial learning rate, $\beta{1}$, $\beta{2}$, weight decay $\lambda$ were set to $2 \times 10^{-4}$, 0.8, 0.99, and 0.01, respectively. Furthermore, we decreased the learning rate by $0.999^{1/8}$ after every epoch.

\subsection{Subjective evaluation by MOS tests}\label{sec:subjective evaluation}
We evaluated the voice quality and expressiveness of the SVS models by MOS and CMOS tests with 30 subjects\footnote{Many previous studies on TTS and SVS listed in our references present the result of MOS test evaluated by 10-20 subjects \cite{UniTTS2021, FastSpeech22020, FastSpeech2019, XiaoiceSing2020, HifiSinger2020}}. The subjects evaluated the voice quality of the synthesized voices in terms of pronunciation accuracy, sound quality, and naturalness. They also evaluated the expressiveness of the models by singer similarity, emotional type similarity, and emotional intensity similarity. To assess singer and emotional type similarities, we asked the subjects to compare the timbre and emotional type of the synthesized samples to those of the ground truth samples. Regarding the similarity in emotional intensity, the subjects first listened to a pair of ground truth samples sung with two different emotional intensity levels, and then listened to a pair of synthesized samples with the same emotional intensity levels as those of the ground truth samples. Finally, the subjects evaluated how close the gap in emotional intensities between the synthesized samples is to the gap between the ground truth samples.

%and remember the difference in recognized emotion between the two samples. Then, we asked the subjects to listen to the synthesized samples of the same emotional intensities as those of the ground-truth samples. Finally, the subjects evaluated how closely the gap in the emotional intensities of the synthesized samples was similar to the gap in the ground-truth samples.
% 감정 강도 변화 차이의 유사도를 평가

\begin{table*}[t!] %% Table of MOS: Naturalness / Singer Similarity
\caption{The results of MOS tests for voice quality and expressiveness}
\label{table:mos-evaluation}
%\vskip 0.1in
\begin{center}
\begin{small}
\begin{sc}
\begin{tabular}{l|ccc|cccr}
\hline
%\toprule
& \multicolumn{3}{c}{\textbf{voice quality}($\uparrow$)} & \multicolumn{3}{c}{\textbf{expressiveness}($\uparrow$)}\\
 & & & & Singer's& Emotional & Emotional\\
 & Pronunciation & Sound & & Timbre & Type & Intensity \\
Methods & Accuracy & Quality & Naturalness & Similarity & Similarity & Similarity \\
\hline
%\midrule
G.T.        & 4.23$\pm$0.30& 4.70$\pm$0.16& 4.71$\pm$0.14 & - & - & - \\
MSME-FFTSinger     & 2.93$\pm$0.34& 2.81$\pm$0.35& 3.18$\pm$0.28 & 3.16$\pm$0.39& 3.35$\pm$0.27 & 3.06$\pm$0.35 \\
MSME-VISinger   & 4.20$\pm$0.27 & 3.72$\pm$0.23& 3.31$\pm$0.30 & 3.00$\pm$0.42& 2.98$\pm$0.28 & 2.35$\pm$0.41 \\
\textbf{Muse-SVS}    & \textbf{4.31$\pm$0.28}& \textbf{4.45$\pm$0.17}& \textbf{4.41$\pm$0.20} & \textbf{3.98$\pm$0.33}& \textbf{4.38$\pm$0.19} & \textbf{4.07$\pm$0.30} \\
\hline
%\bottomrule
\end{tabular}
\end{sc}
\end{small}
\end{center}
\vskip -0.1in
\end{table*}

\begin{table}[t!] %% Table of MOS: Naturalness / Singer Similarity
\caption{The result of ablation studies for voice quality}
\label{table:cmos-evaluation}
%\vskip 0.1in
\begin{center}
\begin{small}
\begin{sc}
\begin{tabular}{lcccr}
\hline
%\toprule
& \multicolumn{2}{c}{\textbf{voice quality}} \\
Methods & Sound Quality & Naturalness \\
\hline\hline
%\midrule
Muse-SVS        & 0& 0 \\
Muse-SVS(w/o SPP)     & +0.03 & -0.53 \\
Muse-SVS(w/o ASPP)   & -0.28& -0.98 \\
Muse-SVS(uncond)    & -0.49& -0.22 \\
\hline
%\bottomrule
\end{tabular}
\end{sc}
\end{small}
\end{center}
%\vskip -0.1in
\end{table}

\subsubsection{\textbf{Evaluation of voice quality and expressiveness}}
Table \ref{table:mos-evaluation} summarizes the evaluation results of Muse-SVS and the baseline models.
Muse-SVS exhibited the highest MOS in all metrics. MSME-VISinger showed the second-best results in the metrics for voice quality evaluation. However, in regard to expressiveness, MSME-FFTSinger exhibited higher MOS than MSME-VISinger.

\subsubsection{\textbf{Ablation study}}
To evaluate the effectiveness of the proposed methods, we conducted a CMOS test for sound quality and naturalness (Table \ref{table:cmos-evaluation}). Removing the statistical pitch variance predictor, as in Muse-SVS(w/o SPP), substantially decreased the naturalness score by -0.53, whereas the sound quality score was slightly increased by +0.03. This implies that the statistical modeling of the F0 frequencies helps synthesize a natural singing voice. When the ASPP-Transformer blocks were replaced by ordinary FFT blocks, as in Muse-SVS(w/o ASPP), the sound quality score decreased by -0.28 while the naturalness score decreased by -0.98. This suggests that ASPP-Transformer, which refers to broad contexts, is effective in improving voice quality, particularly in terms of naturalness. Muse-SVS(uncond) showed the sound quality score decreased by -0.49 and the naturalness score decreased by -0.22. These results suggest that predicting style attribute embeddings conditional on the previous attributes improves sound quality and naturalness.

Table \ref{table:ablation-expressiveness} presents the results of the ablation study for singer similarity, emotional type similarity, and emotional intensity similarity. Muse-SVS exhibited remarkably higher MOS scores than those of other models, suggesting that the proposed methods are effective in improving the expressiveness of the SVS model. Particularly, ASPP-Transformer and statistical pitch predictor significantly improved singer similarity and emotional intensity similarity.

\begin{table}[t!] %% Table of Accuracy: emotional type / Emotional Intensity
\caption{The result of ablation studies for expressiveness}
\label{table:ablation-emotion-similarity}
%\vskip 0.1in
\begin{center}
\begin{small}
\begin{sc}
\begin{tabular}{lcccr}
%\toprule
\hline
& \multicolumn{3}{c}{\textbf{Experssiveness}($\uparrow$)} \\
 & Singer's & Emotional & Emotional  \\
 & Timbre & Type & Intensity  \\
 Methods & Similarity & Similarity & Similarity \\
\hline\hline
%\midrule
%G.T.        & - & - & - \\
\textbf{Muse-SVS}    & \textbf{3.98$\pm$0.33}& \textbf{4.38$\pm$0.19} & \textbf{4.07$\pm$0.30} \\
w/o ASPP  & 2.68$\pm$0.33& 3.69$\pm$0.23 & 3.02$\pm$0.32 \\
w/o SPP    & 3.02$\pm$0.43& 3.88$\pm$0.27 & 2.98$\pm$0.43 \\
uncond     & 3.22$\pm$0.52& 3.57$\pm$0.32 & 3.56$\pm$0.35 \\
\hline
%\bottomrule
\end{tabular}
\end{sc}
\end{small}
\end{center}
%\vskip -0.1in
\label{table:ablation-expressiveness}
\vspace{-0.25in}
\end{table}

\subsection{Quantitative evaluation}
\subsubsection{\textbf{Pitch prediction}}\label{sec:pitch prediction ability}
The emotional SVS should effectively learn the emotions expressed in the training data, which are often conveyed by pitch variances.
To evaluate how closely each SVS model imitates the pitch variances of the training data, we compared the phoneme-wise pitch distributions of the synthesized and ground truth samples by 
Fr\'echet distance \cite{FrechetDistance1957} as in Eq. \ref{eq:pitch prediction error}, where $F[\cdot]$, $L_{n}$, and $N$ denote Fr\'echet distance, the number of phonemes in the $n$-th sample, and the number of samples, respectively. $N(\hat{\mu}_i, \hat{\sigma}_{i}^{2})$, $N(\mu_i,\sigma_{i}^{2})$ are Gaussian distributions parameterized by the mean and variance of the synthesized and ground truth pitch values in the interval of each phoneme, respectively.

%Multi-singer emotional SVS should reflect the changing pitch patterns depending on singer and emotion. To evaluate this, we measured Fr$\acute{e}$chet Distance \cite{FrechetDistance1957} between the distribution of pitch sequence extracted from ground-truth samples and the distribution of pitch sequence extracted from the synthesized samples, as Equation \ref{eq:pitch prediction error}, where $F[\cdot], L_{n}, N$ denote the Fr$\acute{e}$chet Distance, the number of phonemes, the number of samples, and $N(\hat{\mu}_i, \hat{\sigma}_{i}^{2})$, $N(\mu_i,\sigma_{i}^{2})$ are normal distributions parameterized by mean and variance of ground-truth pitch sequence and synthesized pitch sequence corresponding to i-th phoneme, respectively.

\begin{equation} \tag{5}
    Error_p = \frac{1}{N}\sum_{n=1}^{N}\sum_{i=1}^{L_n}F[N(\hat{\mu}_i, \hat{\sigma}_{i}^{2}),N(\mu_i,\sigma_{i}^{2})]
    \label{eq:pitch prediction error}
\end{equation}

We compared three SVS models: MSME-VISinger, Muse-SVS(w/o SPP), and Muse-SVS. MSME-VISinger predicts F0 frequencies at the frame level, whereas the other two models predict F0 frequencies at the phoneme level. MSME-VISinger and Muse-SVS(w/o SPP) predict pitch deterministically, while Muse-SVS predicts statistically. Table \ref{table:pitch prediction performance} presents the evaluation results. The results show that MSME-VISinger and Muse-SVS(w/o SPP) exhibited comparable Fr\'etchel distances. Muse-SVS exhibited significantly lower Fr\'echet distance compared with the two baseline models, indicating that the proposed statistical pitch predictor synthesizes pitch distributions close to those of the training data.

\begin{figure*}[t!]
\centering
    \subfloat[Singer Embedding]{\includegraphics[width=0.3\textwidth, height=5cm]{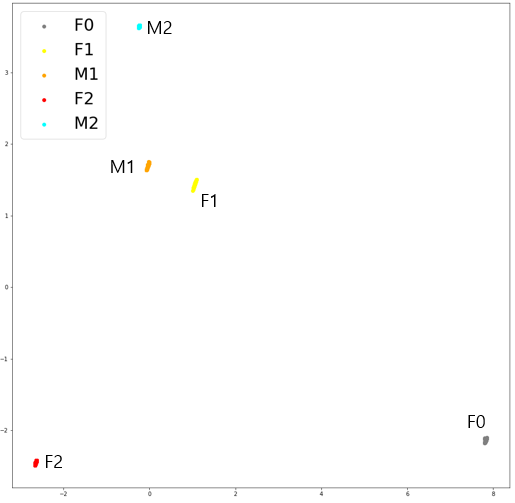}
    \label{figure:singer-embedding}}
    \hfil
    \subfloat[Emotion Embedding]{\includegraphics[width=0.3\textwidth, , height=5cm]{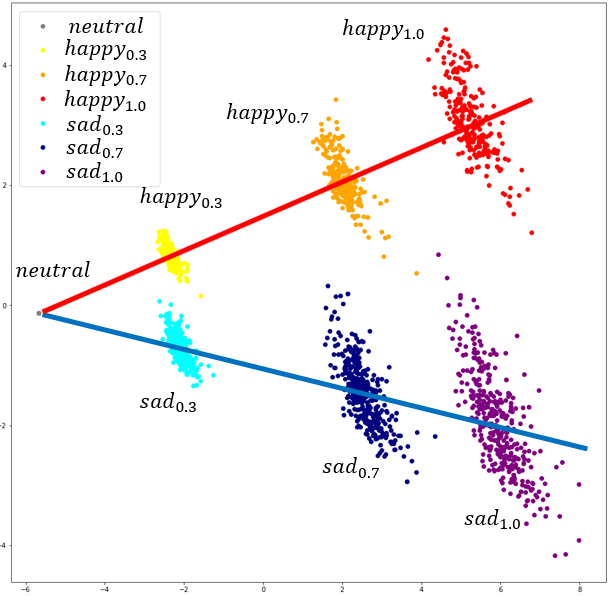}
    \label{figure:emotion-embedding}}
\caption[Visualization of Muse-SVS's residual Singer/Emotion embedding]{The distribution of residual attribute embeddings visualized by PCA. left; residual singer embeddings $R(z_1|y_i)$ colored by singer label, right: residual emotion embeddings $R(z_2|y_i, z_1)$ colored by emotion label. The attribute embeddings are clustered according to their labels, which indicates that Muse-SVS learns attribute embeddings highly correlated with the corresponding attribute labels. We manually drew the lines in the right figure for the convenience of the reader.
}
\label{figure:singer-emotion-embedding}   
\vspace{-0.1in}
\end{figure*}

\begin{table}[t!]
\caption{Fr\'echet distance between pitch distributions (Eq. \ref{eq:pitch prediction error}) in the synthesized and ground truth samples}
\label{table:pitch prediction performance}
%\vskip 0.1in
\begin{center}
\begin{small}
\begin{sc}
\begin{tabular}{lcr}
%\toprule
\hline
Methods & $Error_p (\downarrow)$ \\
\hline\hline
%\midrule
Muse-SVS & \textbf{148.82} \\
Muse-SVS(w/o SPP) & 174.89 \\
MSME-VISinger & 172.64 \\
\hline
%\bottomrule
\end{tabular}
\end{sc}
\end{small}
\end{center}
\vspace{-0.2in}
%\vskip -0.1in
%\label{table:expressiveness}
\end{table}

\subsubsection{\textbf{Synchronization with note duration sequence}}
To evaluate the effectiveness of the proposed CRDP in synchronization with the score, we measured synchronization error for the test samples using Eq. \ref{eq:normalized synchronization error}, where $N$, $L_n$, $\hat{d}_i$, and $\bar{d}_i$ denote the number of samples, the number of phonemes in each sample, the predicted phoneme duration, and the note duration, respectively. Eq. \ref{eq:normalized synchronization error} measures the synchronization error normalized by the total length of the song. In addition, we measured the RMSE between $\hat{d}_i$ and ground truth phoneme duration $d_i$, as Eq. \ref{eq:duration RMSE}.

%We evaluated the effectiveness of the Context-aware Residual Duration Predictor (CRDP) by measuring the cumulative duration prediction error on the test samples as Equation \ref{eq:normalized synchronization error}, where $N$, $L_i$, $\hat{d}$, and $\bar{d}$ denote the number of samples, the number of phonemes in each sample, the duration predicted by the model, and the note duration, respectively. Please note that $Error_d$ measures duration error at the sample-level rather than at the phoneme level.

\begin{equation} \tag{6}
    Error_s \;=\;\frac{1}{N}\sum^{N}\frac{1}{\sum_{i=1}^{L_n}\bar{d}_i}|\sum_{i=1}^{L_n}\hat{d}_i - \sum_{i=1}^{L_n}\bar{d}_i|
    \label{eq:normalized synchronization error}
\end{equation}

\begin{equation} \tag{7}
    RMSE_{d} \;=\;\frac{1}{N}\sum^{N}\sqrt{\frac{1}{L_n}\sum_{i=1}^{L_n}(d_i-\hat{d}_i)^2}
    \label{eq:duration RMSE}
\end{equation}

%''' table start
\begin{table}[h]
\caption{The synchronization and duration prediction performances of duration predictors. 2nd column: the average of synchronization errors at the end of each sample (Eq. \ref{eq:normalized synchronization error}). 3rd column: the RMSE of phoneme-level duration prediction in seconds and the number of frames (Eq. \ref{eq:duration RMSE}).}
\label{table:synchronization error}
%\vskip 0.1in
\begin{center}
\begin{small}
\begin{sc}
\begin{tabular}{lcr}
%\toprule
\hline
Methods & $Error_{s}(\downarrow)$ & $RMSE_{d}(\downarrow)$\\
\hline\hline
%\multirow{2}{*}{Method} & \textbf{Duration} \\
%& \textbf{Error}($\downarrow$) \\
%\midrule
note normalization \cite{VISinger2022}     & 4.7\% & 0.025 (2.33)\\
syllable duration loss \cite{XiaoiceSing2020}  & 6.6\% & 0.027 (2.51)\\
%syllable duration loss \cite{XiaoiceSing2020}  & 4.6\% \\      % 본문에는 6.6%로 명시. 표 오류로 보여서 6.6\%로 수정했음. 아닐 경우 본문과 함께 정정할 것.
CRDP (proposed)    & \textbf{0.2\%} & 0.026 (2.44)\\
\hline
%\bottomrule
\end{tabular}
\end{sc}
\end{small}
\end{center}
%\vskip -0.1in
\label{table:expressiveness}
\vspace{-0.2in}
\end{table}

We compared the proposed CRDP with two baseline duration predictors that apply note normalization \cite{VISinger2022} and syllable duration loss \cite{XiaoiceSing2020}, respectively. The first baseline predictor learns and predicts the ratio of the ground truth phoneme duration to the corresponding note duration, and then estimates phoneme duration by multiplying the corresponding note duration by the predicted ratio. The second baseline predictor was trained by a combination of phoneme-level and syllable-level duration losses \cite{XiaoiceSing2020}, and it directly predicts phoneme duration.

Table \ref{table:synchronization error} presents the evaluation results. The two baseline predictors exhibited synchronization errors of 4.7\% and 6.6\%, respectively. By contrast, the proposed CRDP exhibited a remarkably lower synchronization error of 0.2\%. Regarding phoneme-level duration prediction, the first baseline predictor exhibited the lowest RMSE, while the proposed CRDP exhibited showed the second lowest. However, the differences in phoneme-level RMSEs were insignificant. These results demonstrate that CRDP significantly improves synchronization performance while maintaining phoneme-level prediction accuracy.

%original (교수님)
%Table \ref{table:synchronization error} presents the evaluation results. The two baseline predictors exhibited synchronization errors of 4.7\% and 6.6\%, respectively. In contrast, the proposed CRDP exhibited a synchronization error of 0.2\%, demonstrating its effectiveness in synchronizing with the score.

\subsection{Visualization analysis}
\label{subsubsec:visualization_results}

\begin{figure}[t!]
    \centering
    \includegraphics[width=0.48\textwidth,height=7cm]{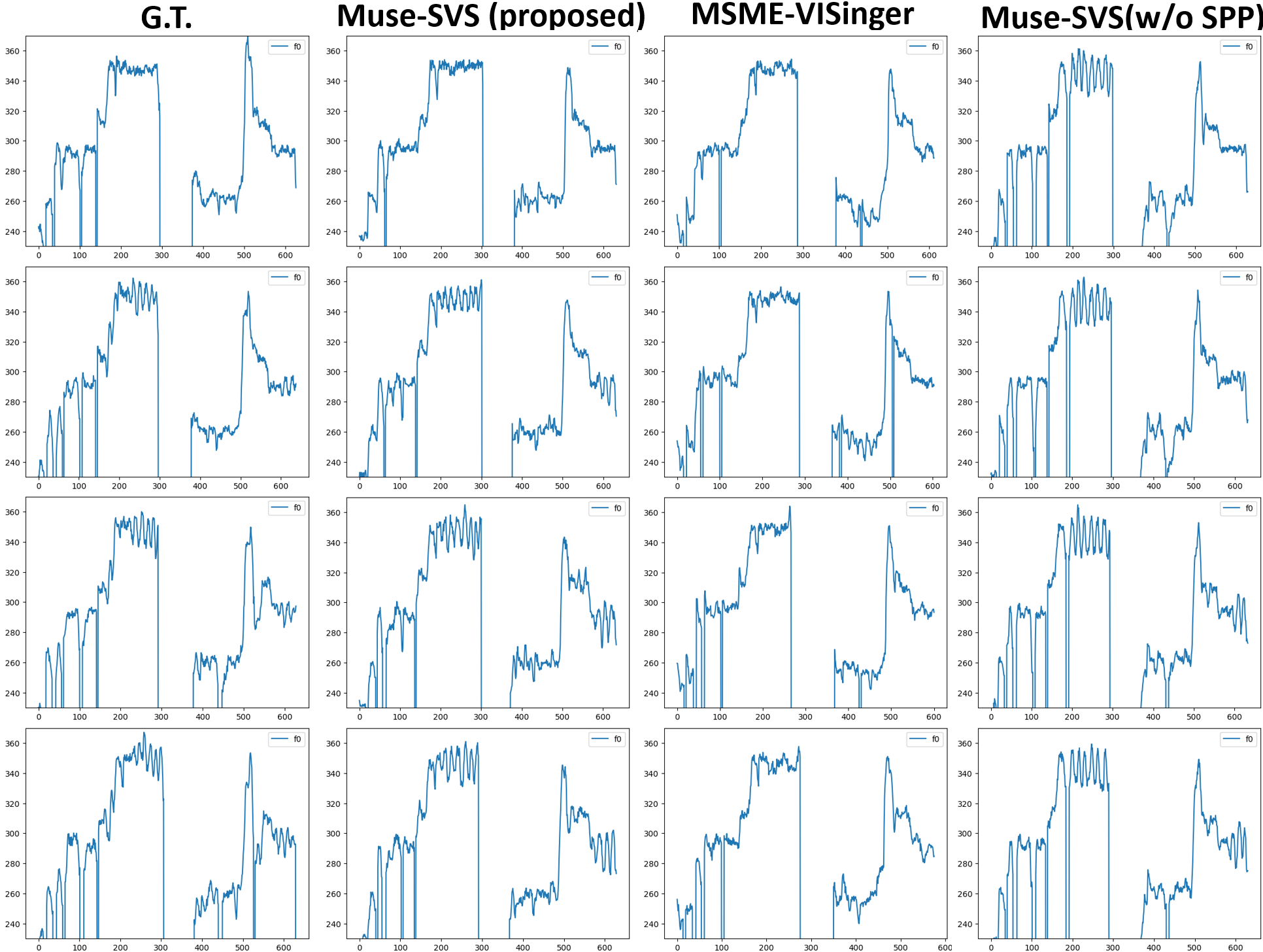}
    \caption{The F0 contours of the singing voices synthesized with different emotional intensities. From the top, each row displays the pitch contours synthesized with emotional intensity levels $neutral$, $sad_{0.3}$, $sad_{0.7}$, and $sad_{1.0}$, respectively.}
    \label{figure:f0-vi-woutSPP-ours}
\end{figure}

\subsubsection{\textbf{Visualization of embedding spaces}}
Muse-SVS represents style attributes $z_k$ by conditional residual embeddings $R(z_k|y_i, z_{\leq k})$ in a unified embedding space. We visualized the distribution of the residual embeddings using principal component analysis (PCA). Fig. \ref{figure:singer-emotion-embedding} displays the distribution of the residual singer and emotion embeddings colored by singer and emotion labels, respectively. The singer and emotion embeddings are clustered according to their labels, indicating that the proposed methods learn attribute embeddings that are highly correlated with the corresponding attribute labels. In particular, due to the embedding interpolation used in training, the embedding distributions of the emotional intensity levels are linearly aligned as shown in Fig. \ref{figure:emotion-embedding}.

\subsubsection{{\textbf{F0 contours varying with emotional intensity}}}
To analyze the effectiveness of the proposed statistical pitch predictor, we visualized the pitch contours synthesized by Muse-SVS and two baseline models, MSME-VISinger and Muse-SVS(w/o SPP), as depicted in Fig. \ref{figure:f0-vi-woutSPP-ours}. Muse-SVS controlled the strength of vibrato according to the emotional intensity, while MSME-VISinger failed to exhibit significant differences according to the emotional intensity. However, when the statistical pitch predictor was replaced by a deterministic pitch predictor, as in Muse-SVS(w/o SPP), the model was unable to control the strength of vibrato according to the emotional intensity anymore.

\subsubsection{{\textbf{Predicted durations varying with emotional intensity}}}

\begin{figure}[t!]
     \vspace{0.1in}
     \centering
     \includegraphics[width=0.5\textwidth,height=6cm]{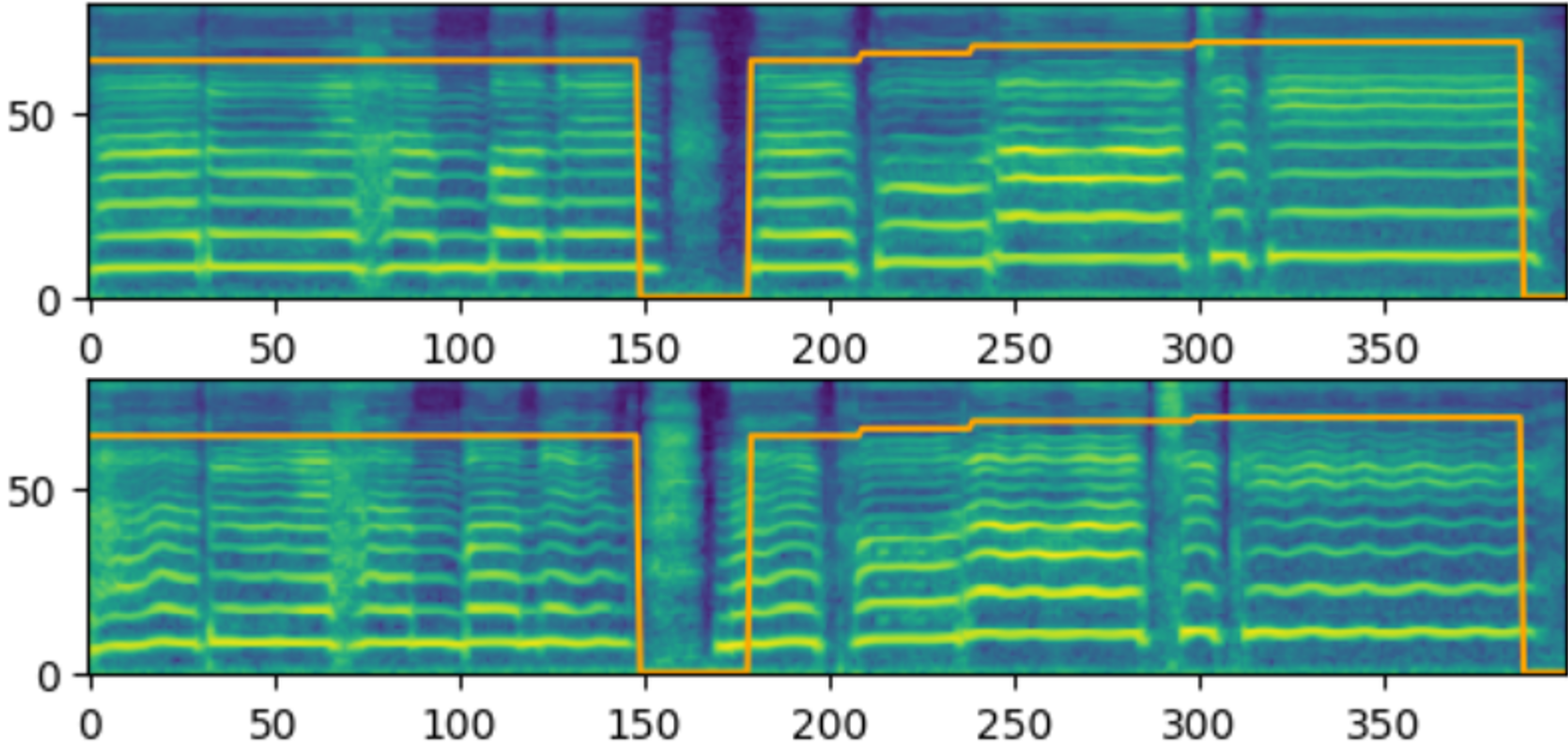}

    \caption{Mel-Spectrograms synthesized by Muse-SVS with emotional intensity levels $neutral$ (top) and $happy_{1.0}$ (bottom). The orange lines denote the note pitch sequence. Muse-SVS predicted phoneme durations differently depending on emotional intensity while maintaining sync with the note pitch sequence until the end of the song.}
     % \caption{Mel-Spectrograms synthesized with different emotional intensities. From the top, each row of the figure exhibits the Mel-Spectrogram synthesized with emotional intensity levels $neutral$, and $happy_{1.0}$. The orange lines exhibit the note pitch sequence.}
     \label{figure:mel-spectrogram}
% %\vskip -0.2in
 \end{figure}

We visualized the Mel-spectrogram synthesized by Muse-SVS with different emotional intensity levels, as shown in Fig. \ref{figure:mel-spectrogram}.
Muse-SVS predicted phoneme durations differently according to emotional intensity. Nevertheless, both Mel-spectrograms remain synchronized with the note pitch sequence until the end of the song.

 \begin{figure}[t!]
    \centering
    \includegraphics[width=0.5\textwidth,height=4.5cm]{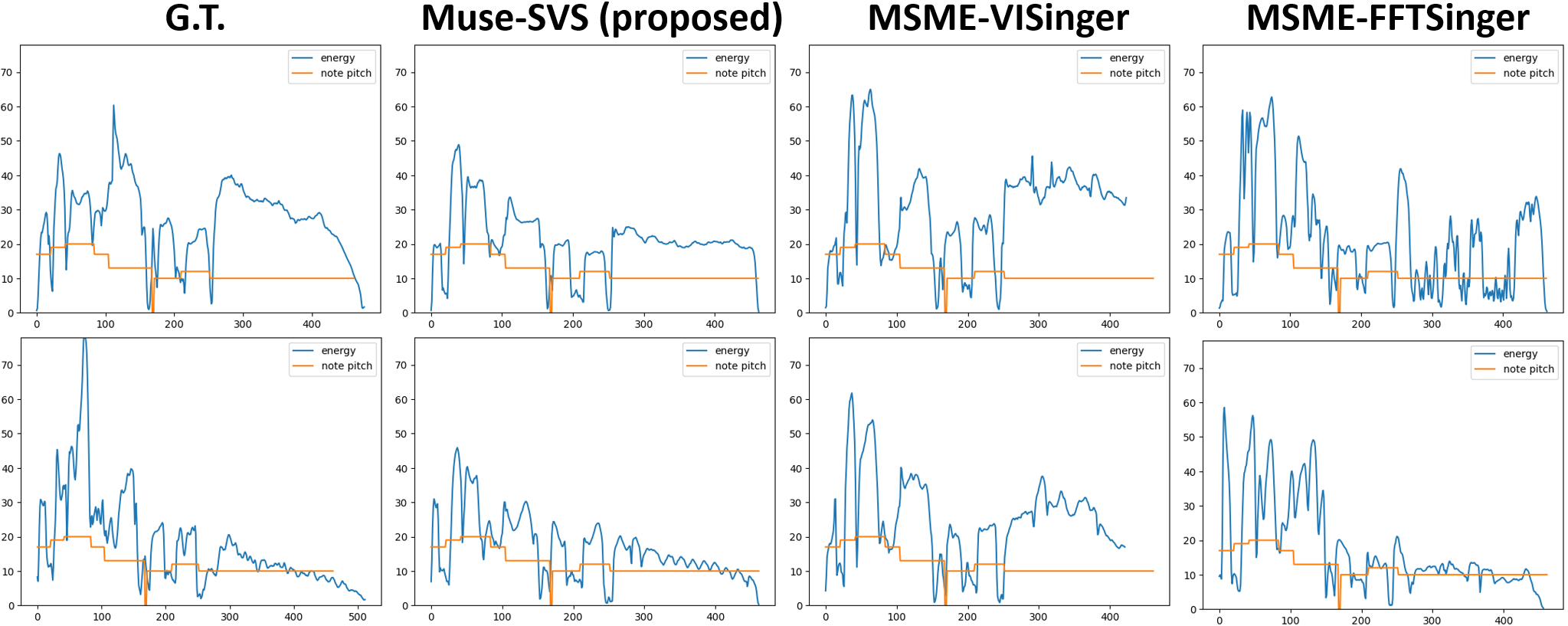}
    \caption{The energy contours of the singing voices synthesized with $neutral$ (top) and $sad_{1.0}$ (bottom). With emotional intensity level $sad_{1.0}$, Muse-SVS produced more fluctuating energy contour and expressed decrescendo and vibrato.}
    \label{figure:energy-contours}
\vspace{-0.2in}
\end{figure}

\subsubsection{{\textbf{Energy contours varying with emotional intensity}}}
% kor version
% 우리는 Muse-SVS가 감정 강도에 따라서 노래의 볼륨(energy)을 다르게 조절한다는 것을 보이기 위해, 서로 다른 감정 강도로 합성한 노래의 energy contour를 시각화 하였다. (Fig. \ref{figure:energy-contours}). 그림의 첫번 째 행은 $neutral$ 로 합성한 음성의 energy contour 들이며, 두번 째 행은 $sad_{1.0}$ 으로 합성한 음성의 energy contour 들이다. 먼저, 그림에서 G.T.의 energy contour들을 보면, 감정 강도가 강해짐에 따라, energy contour의 변동이 많아졌다. 이에 따라, energy contour가 더 뾰족한 peak들을 형성하게 되었다. 또한, 감정 강도가 강해짐에 따라 볼륨이 점점 감소하는 경향을 보였다 (decrecendo). Muse-SVS는 감정 강도가 강해짐에 따라 G.T.에서 나타나는 에너지 변화 경향을 잘 반영하였다. 하지만, VISinger.MSME 와 FFTSinger.MSME는 이러한 변화들을 잘 반영하지 못 했다. 

We visualized the energy contours of the synthesized samples to check whether Muse-SVS controls energy according to emotional intensity. Fig. \ref{figure:energy-contours} displays the energy contours synthesized by Muse-SVS, MSME-VISinger, and MSME-FFTSinger. In the ground truth samples, as the emotional intensity level changes from $neutral$ to $sad_{1.0}$, the energy contour fluctuated more while the decrescendo and vibrato become pronounced. Muse-SVS produced a similar change in energy contour, while the other models did not, as shown in Fig. \ref{figure:energy-contours}.

%To evaluate the model's ability of controlling volume based on emotional intensities, we plotted the energy contours produced by Muse-SVS, VISinger.MSME, and FFTSinger.MSME, as Fig. \ref{figure:energy-contours}. From the top, each row displays the energy contour synthesized with emotional intensity levels $neutral$ and $sad_{1.0}$. When the emotional intensity increases, the contours of G.T. exhibits the following changes: 1)more fluctuations, 2) gradual decrement in energy (decrescendo). Muse-SVS controlled energy, reflecting the changes in the G.T. samples. However, MSME-VISinger and MSME-FFTSinger failed to appropriately display the changes.

\section{Conclusion}
In this study, we proposed Muse-SVS, the first multi-singer emotional singing voice synthesizer that expresses multiple levels of emotional intensities. Muse-SVS synthesizes singing voices from lyrics, note pitch, and note duration controlling multiple attributes such as singer ID and emotional intensity. It synthesizes subtle variations in pitch, energy, and phoneme duration according to emotional intensity while accurately synchronizing with the music score. To avoid interference between non-hierarchically correlated attributes, Muse-SVS represents multiple style attributes by a joint embedding in a unified embedding space that encodes all attributes and their relations together. We presented multiple novel techniques to improve the voice quality and expressiveness of SVS, including a statistical pitch predictor, context-aware duration predictor, and ASPP-Transformer. Compared with the baseline models, Muse-SVS exhibited improved voice quality, expressiveness, and synchronization performance in MOS tests and quantitative evaluations. We also presented visualization results demonstrating that Muse-SVS learns attribute embeddings highly correlated with the corresponding attribute labels, and that Muse-SVS successfully controls subtle changes in pitch, energy, and phoneme duration according to emotions.

%In this paper, we proposed Muse-SVS, the first multi-singer emotional singing voice synthesizer that controls emotional intensity based on the unified embedding space. Muse-SVS synthesizes singing voices from lyric, note pitch, and note duration controlling multiple attributes such as singer ID and emotional intensity. It represents emotional intensity by controlling the fine variation of pitch, energy, and phoneme duration. Muse-SVS applies such novel techniques as Statistical Pitch Predictor, Context-aware Residual Duration predictor, and ASPP-Transformer to accurately predict and reflect multiple attributes according to singer ID and emotional intensities. Additionally, we proposed novel emotion interpolation and extrapolation techniques. In experiments, Muse-SVS synthesized singing voices reflecting the specified emotional intensity level. We also presented multiple quantitative evaluation results and visualization results that confirm the effectiveness of the proposed method in expressing the type and intensity of emotions.

\bibliographystyle{IEEEtran}
\bibliography{references}

\begin{IEEEbiography}[{\includegraphics[width=1in,height=1.25in,clip,keepaspectratio]{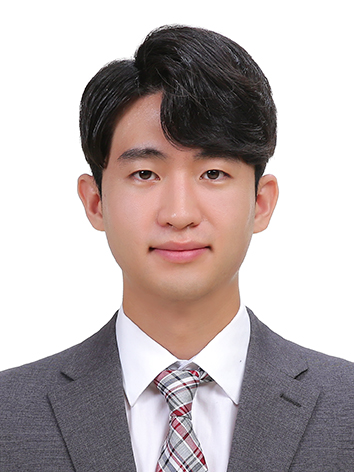}}]{Sungjae Kim}
received the B.S. and M.S. degree in CSEE (Computer Science and Electrical Engineering) at Handong Global University. He is currently a Ph.D student in CSEE at Handong Global University. Since 2019, he is a student researcher at DL-LAB of Handong Global University, working under the supervision of Prof. Injung Kim. His research interest includes deep learning, speech synthesis, singing voice synthesis, and speech recognition.
\end{IEEEbiography}

\begin{IEEEbiography}[{\includegraphics[width=1in,height=1.25in,clip,keepaspectratio]{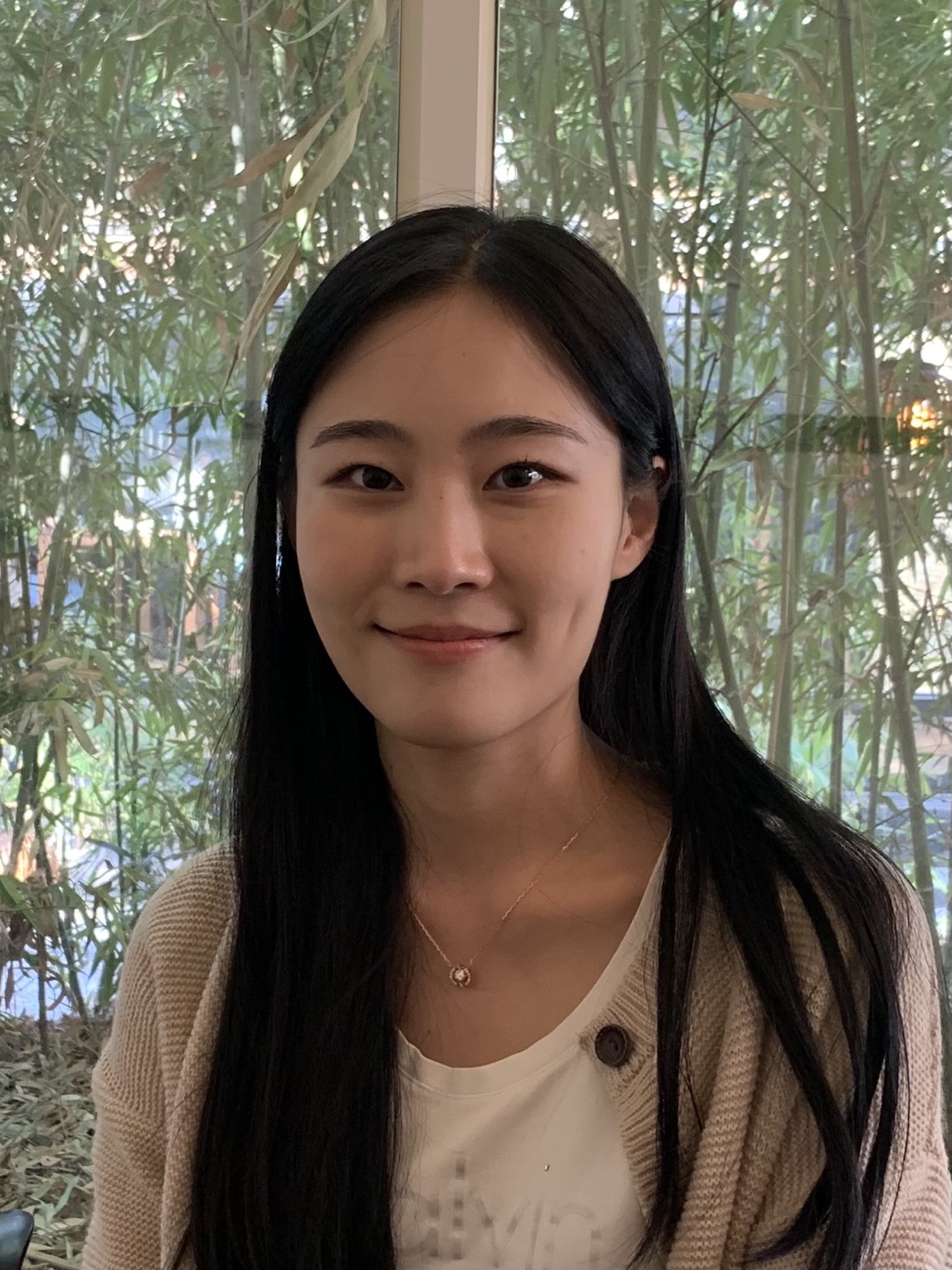}}]{Yewon Kim}
received the B.S. degree in CSEE at Handong Global University. She is currently a M.S. student in CSEE at Handong Global University. From 2020 to June 2022, she worked as a student researcher at HAIL(Handong AI Lab) of Handong Global University with Prof. Charmgil Hong, participating in several industry projects as a student researcher. Since July 2022, she is a student researcher at DL-LAB of Handong Global University, working under the supervision of Prof. Injung Kim, participating projects related to speech synthesis and singing voice synthesis. Her research interest includes speech synthesis, singing voice synthesis and deep learning.
\end{IEEEbiography}

\begin{IEEEbiography}[{\includegraphics[width=1in,height=1.25in,clip,keepaspectratio]{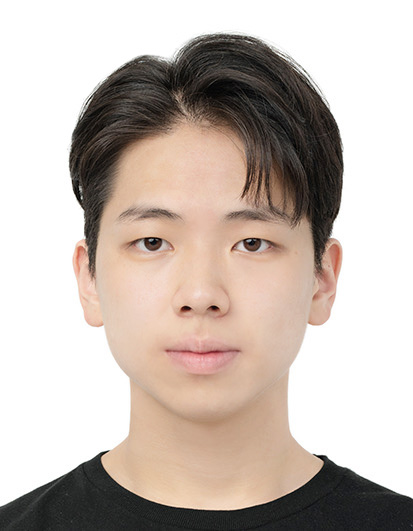}}]{Jewoo Jun}
is currently a B.S. student in CSEE at Handong Global University. From 2022, he is a student researcher at DL-LAB of Handong Global University, working under the supervision of Prof. Injung Kim. His research interests include speech synthesis and singing voice synthesis.
\end{IEEEbiography}

\begin{IEEEbiography}[{\includegraphics[width=1in,height=1.25in,clip,keepaspectratio]{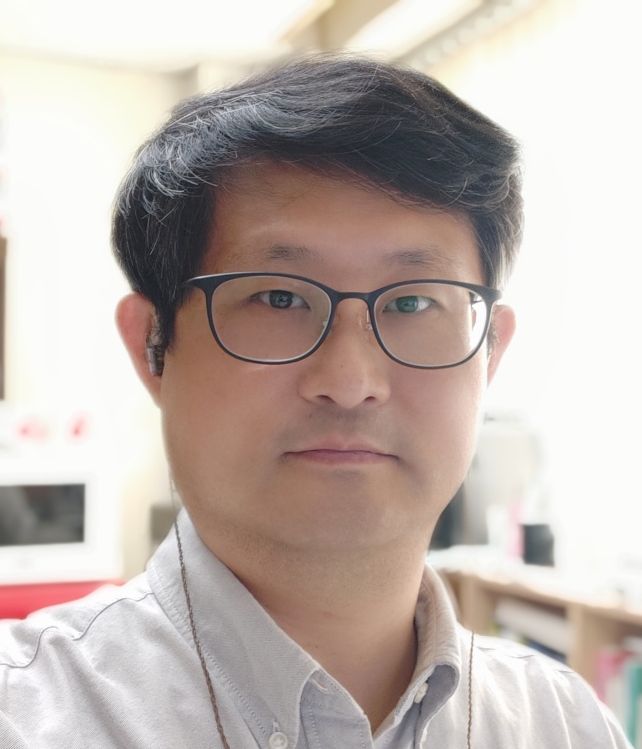}}]{Injung Kim}
is a professor of CSEE, Handong Global University since 2006.
He received B.S., M.S., and Ph.D. degrees in Computer Science from KAIST (Korea Advanced Institute of Science and Technology). He was a senior research engineer of Inzisoft. He was the Head of the School of CSEE, Program Director of the CS major, and an AI advisor of Samsung SW Center and POSCO, and currently, he is research advisor of multiple AI companies. His research interests include deep learning, image analysis and synthesis, speech synthesis, data analysis and prediction, recommendation system, outlier detection, and natural language processing.
\end{IEEEbiography}

%\vspace{11pt}

%\bf{If you will not include a photo:}\vspace{-33pt}
%\begin{IEEEbiographynophoto}{John Doe}
%Use $\backslash${\tt{begin\{IEEEbiographynophoto\}}} and the author name as the argument followed by the biography text.
%\end{IEEEbiographynophoto}

\vfill

\end{document}